\documentclass[twocolumn]{aastex63}

\usepackage{multirow}
\usepackage{color}
\received{March 16, 2020}
\revised{April 27, 2020}
\accepted{14 May, 2020}

\shorttitle{Inner disk of RY~Tau}
\shortauthors{Davies et al.}

\begin{document}

\title{The inner disk of RY~Tau: evidence of stellar occultation by the disk atmosphere at the sublimation rim from $K$-band continuum interferometry}

\correspondingauthor{Claire L.\ Davies}
\email{c.davies3@exeter.ac.uk}

\author[0000-0001-9764-2357]{Claire L. Davies}
\affiliation{Astrophysics Group, School of Physics, University of Exeter, Stocker Road, Exeter, EX4 4QL, UK}

\author[0000-0001-6017-8773]{Stefan Kraus}   
\affiliation{Astrophysics Group, School of Physics, University of Exeter, Stocker Road, Exeter, EX4 4QL, UK}

\author[0000-0001-8228-9503]{Tim J.\ Harries}   
\affiliation{Astrophysics Group, School of Physics, University of Exeter, Stocker Road, Exeter, EX4 4QL, UK}

\author[0000-0002-3380-3307]{John D.\ Monnier}  
\affiliation{Department of Astronomy, University of Michigan, Ann Arbor, MI 48109, USA}

\author[0000-0003-0350-5453]{Brian Kloppenborg} 
\affiliation{Department of Physics and Astronomy, Georgia State University, Atlanta, GA 30302, USA}

\author[0000-0002-1327-9659]{Alicia Aarnio} 
\affiliation{University of Colorado Boulder, Boulder, CO 80303, USA}

\author[0000-0002-8376-8941]{Fabien Baron}    
\affiliation{Department of Physics and Astronomy, Georgia State University, Atlanta, GA 30302, USA}

\author[0000-0002-2144-0991]{Rebeca Garcia Lopez}
\affiliation{School of Physics, University College Dublin, Belfield, Dublin 4, Ireland}

\author[0000-0003-0447-5866]{Rafael Millan-Gabet}
\affiliation{Infrared Processing and Analysis Center, California Institute of Technology, Pasadena, CA, 91125, USA}

\author{Robert Parks} 
\affiliation{Department of Physics and Astronomy, Georgia State University, Atlanta, GA 30302, USA}

\author{Ettore Pedretti}  
\affiliation{STFC Rutherford Appleton Laboratory, Harwell Science \& Innovation Campus, OX11 0QX, UK}

\author[0000-0003-3099-757X]{Karine Perraut}
\affiliation{Univ. Grenoble Alpes, CNRS, IPAG, 38000 Grenoble, France}

\author{Judit Sturmann}  
\affiliation{The CHARA Array of Georgia State University, Mount Wilson Observatory, Mount Wilson, CA 91203, USA}

\author{Laszlo Sturmann} 
\affiliation{The CHARA Array of Georgia State University, Mount Wilson Observatory, Mount Wilson, CA 91203, USA}

\author[0000-0002-0114-7915]{Theo A.\ ten Brummelaar} 
\affiliation{The CHARA Array of Georgia State University, Mount Wilson Observatory, Mount Wilson, CA 91203, USA}

\author{Yamina Touhami} 
\affiliation{Department of Physics and Astronomy, Georgia State University, Atlanta, GA 30302, USA}


\begin{abstract}
We present models of the inner region of the circumstellar disk of RY~Tau which aim to explain our near-infrared ($K$-band: $2.1\,\mu$m) interferometric observations while remaining consistent with the optical to near-infrared portions of the spectral energy distribution. 
Our sub-milliarcsecond resolution CHARA Array observations are supplemented with shorter baseline, archival data from PTI, KI and VLTI/GRAVITY and modeled using an axisymmetric Monte Carlo radiative transfer code. 
The $K$-band visibilities are well-fit by models incorporating a central star illuminating a disk with an inner edge shaped by dust sublimation at $0.210\pm0.005\,$au, assuming a viewing geometry adopted from millimeter interferometry ($65^{\circ}$ inclined with a disk major axis position angle of $23^{\circ}$). This sublimation radius is consistent with that expected of Silicate grains with a maximum size of $0.36-0.40\,\mu$m contributing to the opacity and is an order of magnitude further from the star than the theoretical magnetospheric truncation radius. The visibilities on the longest baselines probed by CHARA indicate that we lack a clear line-of-sight to the stellar photosphere. Instead, our analysis shows that the central star is occulted by the disk surface layers close to the sublimation rim. While we do not see direct evidence of temporal variability in our multi-epoch CHARA observations, we suggest the aperiodic photometric variability of RY~Tau is likely related temporal and/or azimuthal variations in the structure of the disk surface layers.
\end{abstract}

 

\section{Introduction}
The reprocessing of starlight by dust in the innermost regions of the disks of young stellar objects (YSOs) produces strong near-infrared (NIR) continuum emission in excess of that expected from purely photospheric emission. The milliarcsecond (mas) and sub-mas resolution provided by NIR interferometry at $\sim1-3\,\mu$m can be used to spatially resolve this region and shed light on the shape and structure of the environments in which planets form and evolve. The earliest NIR interferometric studies of disks showed that dust had a finite inner limit and did not extend down to the stellar surface \citep{Millan99, Akeson00}. The location of this inner edge is likely forged by dust sublimation \citep{Tuthill01, Monnier02} with the slope of the inner edge size--stellar luminosity relation indicating a dust sublimation temperature, $T_{\rm{sub}}\sim1800\,$K \citep{Lazareff17, Gravity19}. 

The lack of any strong viewing-angle dependency to the closure phase signals, $\phi_{\rm{CP}}$, obtained via NIR interferometry further indicated that this sublimation rim was likely a curved surface rather than a vertical wall \citep{Monnier05}. This curvature arises naturally due to the dependence of $T_{\rm{sub}}$ and the grain cooling efficiency on the gas density, the size distribution of dust grains, the preferential settling of larger grains toward the disk midplane, and the relative abundance of different grain compositions \citep{Pollack94, Isella05, Tannirkulam07, Kama09, McClure13}.

Herein, we focus on \object{RY~Tau} (spectral type G1, \citealt{Calvet04}) and study the shape and structure of its circumstellar NIR-emitting region. The existence of circumstellar material around \object{RY~Tau} was first identified through its strong infrared (IR) excess \citep{Mendoza68}. Analysis of \object{RY~Tau}'s spectral energy distribution (SED) across IR wavelengths led to its classification as a pre-transitional disk \citep{Marsh92, Furlan09, Espaillat11}: the NIR excess is typical of accretion disks but the relative dearth of mid-IR (MIR) excess flux indicates the likely presence of a dust cavity or optically thin region of the disk. A dust cavity was indeed observed via high-resolution millimeter (mm) imaging obtained with the Combined Array for Research in Millimeter-wave Astronomy (CARMA, \citealt{Isella10}) and the Atacama Large Millimeter Array (ALMA, \citealt{Long18, Long19}). The object's micro-jet emission, observed at optical \citep{StOnge08} and NIR wavelengths \citep{Garufi19}, and its relatively strong mass accretion rate, (typical of disks with substantial mass reservoirs in their innermost disk regions; \citealt{Calvet04, Mendigutia11}) also support this classification.

Direct observation of the inner tens of au of the disk has remained difficult. Hubble Space Telescope (HST, \citealt{Agra09}) and polarized intensity images obtained in the optical and NIR with VLT/SPHERE \citep{Garufi19} and Subaru/HiCIAO \citep{Takami13} are dominated by an optically thin scattering layer above the disk surface. The astrophysical nature of this scattering surface remains unclear with a remnant spherical envelope or a dusty outflow caused by a magnetospheric or photo-evaporative wind providing possible explanations. 

NIR and MIR interferometric observations of \object{RY~Tau} -- obtained with the Palomar Testbed Interferometer (PTI, \citealt{Akeson05}), the Infrared Optical Telescope Array (IOTA, \citealt{Monnier05}) and the Very Large telescope Interferometer's MID-infrared Interferometer instrument (VLTI/MIDI, \citealt{Schegerer08}) -- have previously probed the circumstellar emission on sub-au to au scales. However, these observations have been limited by (i) the $\lesssim100\,$m maximum baseline lengths of the interferometric arrays; (ii) the poor baseline position angle, PA$_{b}$, coverage of the observations; (iii) poor constraints on the exact circumstellar-versus-stellar flux contribution due to the intrinsically variable nature of \object{RY~Tau}. \citet{Akeson05} and \citet{Monnier06} attempted to estimate the characteristic size of the NIR-emitting region, with model-dependent estimates of $\sim0.2-0.6\,$au (using a stellar distance, $d=140\,$pc), broadly consistent with the expected dust sublimation radius, $R_{\rm{sub}}$, given the object's luminosity ($\sim6-12\,\rm{L_{\odot}}$, e.g. \citealt{Calvet04, Garufi19, Long19}).

In these prior NIR and MIR interferometric studies, the disk inclination, $i_{\rm{d}}$, was either assumed to be face-on \citep[i.e. $i_{\rm{d}}=0^\circ$,][]{Monnier06, Schegerer08}, or left free in the fitting and loosely constrained around $i_{\rm{d}}\approx20-25^{\circ}$ \citep{Akeson05}. This is in stark contrast to the highly inclined ($i_{\rm{d}}\sim60-70^{\circ}$) disk observed by CARMA \citep{Isella10} and ALMA \citep{Pinilla18, Long18, Long19}. A more highly inclined inner disk is also supported by the nature of the optical and IR photometric variability exhibited by \object{RY~Tau} which is likely to arise due to line of sight occultation of the stellar photosphere by circumstellar material \citep{Grankin07, Petrov19}. 

This study continues our analysis of YSOs observed with the Center for High Angular Resolution Astronomy (CHARA) Array's two-telescope (CLASSIC) and three-telescope (CLIMB) combiners (c.f. \citealt{Davies18}, \citealt{Setterholm18}, and \citealt{Labdon19}). A description of our $K$-band observations of \object{RY~Tau} with CLASSIC and CLIMB is presented in Section~\ref{sec:obs}. The $\sim330\,$m maximum baselines of the CHARA Array provide us with unrivaled spatial resolution in the NIR. We supplement our CLASSIC and CLIMB observations with archival short-baseline $K$-band interferometric data (Section~\ref{sec:comple_obs}), thus benefiting from a greatly improved PA$_{b}$ coverage compared to the \citet{Akeson05} and \citet{Monnier05} studies. We build on work conducted by \citet{Tannirkulam08}, \citet{Davies18} and \citet{Labdon19} and use the TORUS Monte Carlo radiative transfer code \citep{Harries19} to explore the shape and structure of the NIR circumstellar emission component. We provide details of our modeling and results in Section~\ref{sec:mod_results} and present a discussion of our results in Section~\ref{sec:discussion}.

\section{Observations and supplementary archival data}
\begin{figure}[t]
  \centering
  \includegraphics[width=0.4\textwidth]{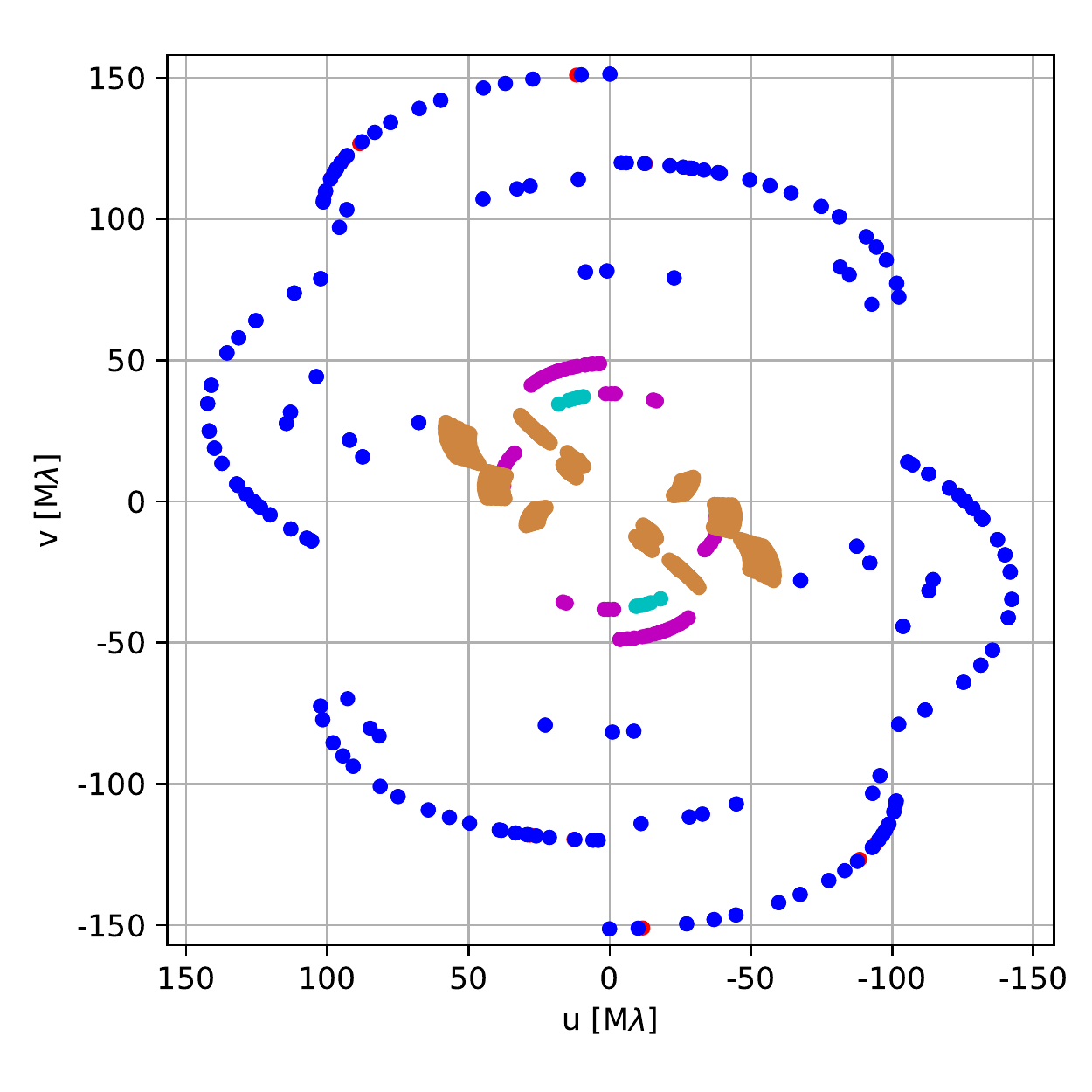}
  \caption{($u, v$)-plane coverage of the compiled $K$-band interferometry. North is up; East is left. CLASSIC and CLIMB observations (see Table~\ref{tab:obslog}) are indicated by red and blue data points, respectively. The supplementary short baseline interferometric data from KI (cyan), PTI (magenta) and VLTI/GRAVITY (orange; see Table~\ref{tab:supldata}) are also shown.}
  \label{fig:uvplane}
\end{figure}

\subsection{CHARA interferometry}\label{sec:obs}
The CLASSIC and CLIMB beam combiners \citep{Brummelaar13} of the CHARA Array were used to obtain $K$-band interferometric observations of \object{RY~Tau} between 2009\,Oct and 2012\,Nov. The CHARA Array is Y-shaped and comprises six $1\,$m class telescopes located at Mount Wilson Observatory with operational baselines of $34$-$331\,$m (corresponding to a maximum resolution\footnote{$\lambda/2B$ with $\lambda$ the operational wavelength ($2.13\,\mu$m) and $B$ the separation between telescopes.} of $0.66\,$mas) \citep{Brummelaar05}. A log of our observations is presented in Table~\ref{tab:obslog}. The ($u, v$)-plane coverage is displayed by the red and blue data points in Fig.~\ref{fig:uvplane}. 

The data were reduced using a pipeline developed at the University of Michigan which is better suited to recovering faint fringes for low visibility data than the standard CHARA reduction pipeline of \citet{Brummelaar12}. Further details regarding the reduction procedure are given in \citet{Davies18}. Calibrator stars were observed before and/or after each observation and used to calibrate the squared visibilities and $\phi_{\rm{CP}}$. None of the calibrators used are known members of binary or multiple systems. Where CLIMB data was obtained for a calibrator, the $\phi_{\rm{CP}}$ were inspected as a further check for binarity. No evidence for the presence of companions (non-zero $\phi_{\rm{CP}}$) were found. Calibrator uniform disk (UD) diameters, retrieved from JMMC SearchCal \citep{Bonneau06, Bonneau11} when available, or from getCal\footnote{http://nexsci.caltech.edu/software/getCal/}, (see Table~\ref{tab:obslog}), were used to calculate the transfer function and are listed in the footnote to Table~\ref{tab:obslog}. The calibrated data will be made accessible in OIFITS format \citep{Pauls05, Duvert17} through the Optical interferometry Database (OiDb; \citealt{Haubois14}) of the JMMC and through the CHARA archive (J. Jones et al. 2020, in preparation), hosted by Georgia State University, following publication. 

\begin{deluxetable}{cccc}
\tabletypesize{\scriptsize}
\tablecaption{CHARA Observation Log\label{tab:obslog}}
\tablehead{
 \colhead{Date} & \colhead{Beam} & \colhead{Stations} & \colhead{Calibrator(s)}\\
 \colhead{(UT)} & \colhead{Combiner}  & \colhead{} & \colhead{}
 }
\startdata
2009 Oct 31    & CLASSIC    & E1 S1   & 1\\
2009 Nov 01    & CLASSIC    & E1 S1   & 2\\
2009 Nov 24    & CLASSIC    & S1 W1   & 3\\
\hline
2010 Sep 29    & CLIMB      & S1 E1 W1 & 2, 4\\
2010 Oct 02    & CLIMB      & S1 E1 W1 & 2, 4\\
2010 Oct 04    & CLIMB      & S1 E1 W1 & 2, 4\\
2010 Dec 02    & CLIMB      & S2 E1 W2 & 1, 2\\
2011 Oct 27    & CLIMB      & S2 E2 W2 & 2\\
2011 Dec 22    & CLIMB      & S2 E2 W1 & 2, 5\\
2012 Nov 26    & CLIMB      & S1 E1 W1 & 2, 6\\
2012 Nov 27    & CLIMB      & S1 E1 W1 & 2
\enddata
\tablecomments{Calibrators and their UD diameters in mas: 1: HD~32480 ($0.221\pm0.016$); 2: HD~24365 ($0.317\pm0.022$); 3: HD~28447 ($0.503\pm0.035$); 4: HD~25461 ($0.245\pm0.017$); 5: HD~30912 ($0.44\pm0.10$); 6: HD~33252 ($0.299\pm0.021$).} 
\end{deluxetable}

\subsection{Complementary short-baseline interferometry}\label{sec:comple_obs}
To probe more extended components of the circumstellar emission from \object{RY~Tau}, we supplemented our CHARA observations with shorter baseline, $K$-band archival interferometric observations (see Table~\ref{tab:supldata}). Calibrated PTI \citep{Colavita99} data, originally published in \citet{Akeson05}, were provided by Rachel Akeson while reduced Keck Interferometer (KI, \citealt{Colavita13}) data were retrieved from the Keck Observatory Archive. The wide-band KI data were calibrated using the NExScI Wide-band Interferometric Visibility Calibration (wbCalib v1.4.4) tool with the flux bias correction and ratio correction options selected. 

Data obtained using the GRAVITY instrument \citep{gravity} of the VLTI were also retrieved from the European Southern Observatory archive. The data were reduced and calibrated using GRAVITY pipeline version 1.1.2 with default settings. We restrict our analysis to the low spectral dispersion ($R\sim30$) GRAVITY fringe tracker data which provides five wavelength channels across the $K$-band. We exclude the first spectral channel from our analysis as these are systematically lower than the other channels (likely due to corruption by the metrology laser which operates at $\lambda=1.08\,\mu$m). The calibrators (and their UDs) used to calibrate the KI and VLTI/GRAVITY data are provided in the footnote to Table~\ref{tab:supldata}. 

\begin{deluxetable}{cccc}
\tabletypesize{\scriptsize}
\tablecaption{Supplementary Interferometric Data\label{tab:supldata}}
\tablehead{ 
 \colhead{Date} & \colhead{Program} & \colhead{Stations} & \colhead{Calibrator(s)}\\
 \colhead{(UT)} & \colhead{ ID} & \colhead{} & \colhead{} 
 }
\startdata
\multicolumn{4}{c}{PTI}\\
\tableline
2001 Sep 24 & --         & NW    & --\\
2001 Sep 27 & --         & NW    & --\\
2001 Oct 03 & --         & NS    & --\\
2001 Oct 17 & --         & NW    & --\\
2001 Nov 07 & --         & NS    & --\\
2001 Nov 17 & --         & NS    & --\\
2001 Nov 22 & --         & NS    & --\\
2003 Oct 14 & --         & SW    & --\\
2003 Oct 15 & --         & SW    & --\\
\tableline
\multicolumn{4}{c}{KI}\\
\tableline
2006 Nov 12 & 32         & K1K2    & 1,2\\
2008 Dec 15 & 48         & K1K2    & 3\\
2010 Nov 24 & 51         & K1K2    & 1\\
\tableline
\multicolumn{4}{c}{VLTI/GRAVITY}\\
\tableline
2017 Dec 10 & 0100.C-0278 & UT1-UT2-UT3-UT4 & 4,5
\enddata
\tablecomments{Calibrators are listed in column 4 when data were (re-)reduced. Their identifiers (and UD diameters in mas) are: 1: HD~27777 ($0.17\pm0.01$); 2: HD~31592 ($0.19\pm0.01$); 3: HD~283934 ($0.071\pm0.014$); 4: HD~58923 ($0.433\pm0.002$); 5: HD~96113 ($0.367\pm0.001$).}
\end{deluxetable}

\subsection{Multi-band photometry and MIR spectroscopy}
Multi-wavelength photometry for \object{RY~Tau} was retrieved from the literature. These data were primarily acquired as an additional assessment of the NIR flux provided by our models. This is vital as visibility modeling is known to be affected by degeneracies between the stellar-to-circumstellar flux contrast and the characteristic size of the emitting region \citep[e.g.][]{Lazareff17}. The collated data is presented in Appendix~\ref{apen:phot} and shown in Fig.~\ref{fig:sed_stellar_direct020} compared to the \citet{Kurucz79} spectrum of a star with effective temperature, $T_{\rm{eff}}=5945\,$K, luminosity, $L_{\star}=11.6\,\rm{L_{\odot}}$, and surface gravity, $\log(g)=3.8$ (see Table~\ref{tab:starParam}). The strong IR excess arising from the presence of circumstellar material is clearly visible. 

\begin{figure}
    \centering
    \includegraphics[width=0.47\textwidth]{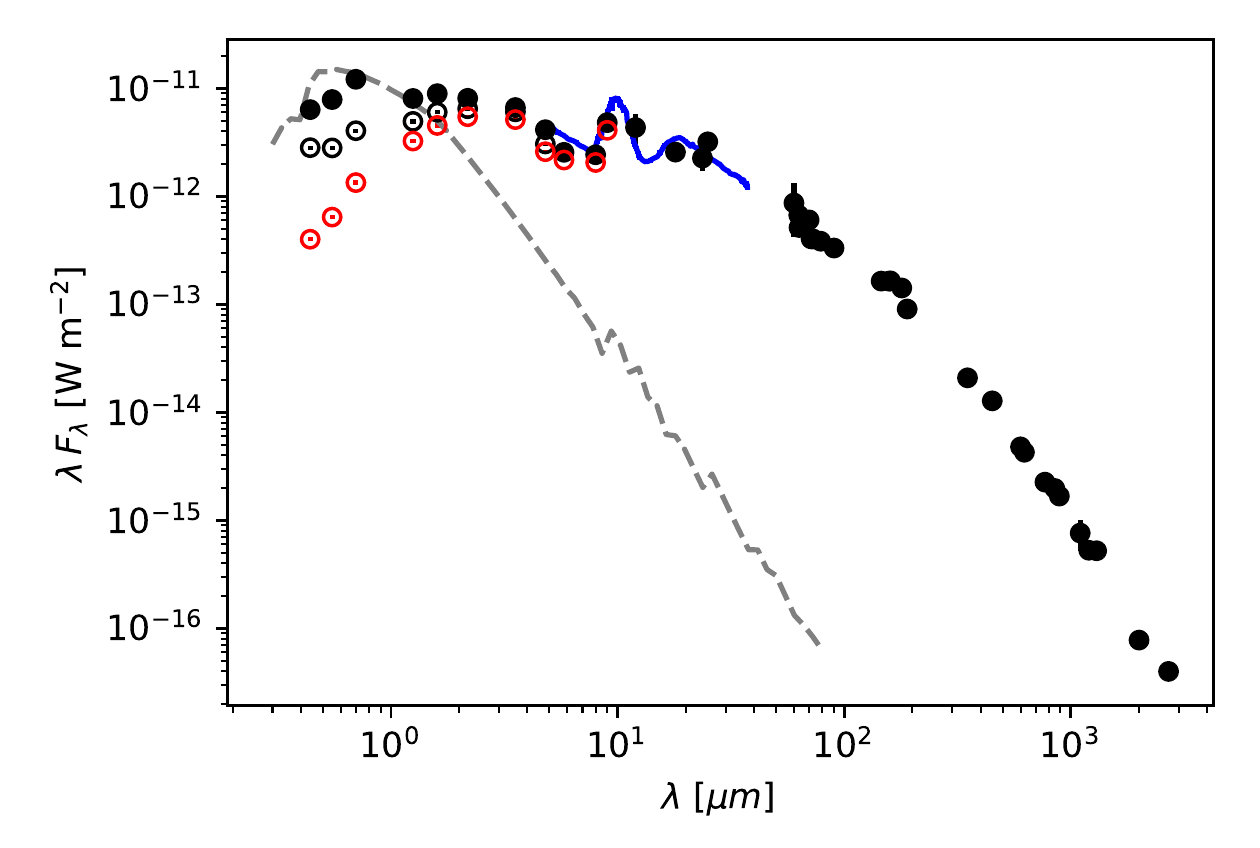}
    \caption{Comparison of our compiled SED for \object{RY~Tau} (see Section~\ref{sec:starParam} and Appendix~\ref{apen:phot} for details) with the spectrum of a star with properties given in Table~\ref{tab:starParam} (grey dashed line). Red and black open circle symbols represent the ``faint'' epoch photometry from \citet{Petrov19} with $A_{\rm{V}}=0$ (i.e. no de-reddening applied) and $A_{\rm{V}}=1.6$, respectively. Black filled circle symbols represent the ``bright'' epoch photometry from \citet{Petrov19} with $A_{\rm{V}}=1.6$. The blue line represents the \emph{Spitzer} spectrum.}
    \label{fig:sed_stellar_direct020}
\end{figure}

As \object{RY~Tau} is variable across optical and NIR wavelengths (e.g. \citealt{Grankin07, Petrov19}), two sets of Johnson-\textit{BVRJHKLM} photometry are tabulated in Appendix~\ref{apen:phot} and shown in Fig.~\ref{fig:sed_stellar_direct020}. These are taken from the \citet{Petrov19} photometric monitoring study and are characteristic of a ``bright'' (black filled circles) and a ``faint'' (red and black open circles) epoch, obtained on 1989\,Oct\,25 and 2016\,Nov\,11, respectively. The red open circles have not been de-reddened (i.e. assumes interstellar extinction, $A_{\rm{V}}=0.0$) while the black open and filled circles have been de-reddened using $A_{\rm{V}}=1.6$ (see Table~\ref{tab:starParam}). As our interferometric data were obtained over several years and the photometry was not obtained contemporaneously with the interferometry, we adopt the red and black filled data as indicators of the upper and lower bounds to the optical and NIR flux allowed in our models.

A post-processed, flux-calibrated \textit{Spitzer} Infrared Spectrograph \citep[IRS;][]{Houck04} spectrum for \object{RY~Tau} \citep[][AORkey 27185920]{Lebouteiller11} was retrieved from the Cornell Atlas of Spitzer/IRS Sources (CASSIS\footnote{The Cornell Atlas of Spitzer/IRS Sources (CASSIS) is a product of the Infrared Science Center at Cornell University, supported by NASA and JPL.} version $7$). This is shown by the blue line in Fig.~\ref{fig:sed_stellar_direct020}. 

\section{Modeling and results}\label{sec:mod_results}
\begin{figure}
    \centering
    \includegraphics[trim=0cm 0.3cm 0cm 0.3cm, clip=true, width=0.45\textwidth]{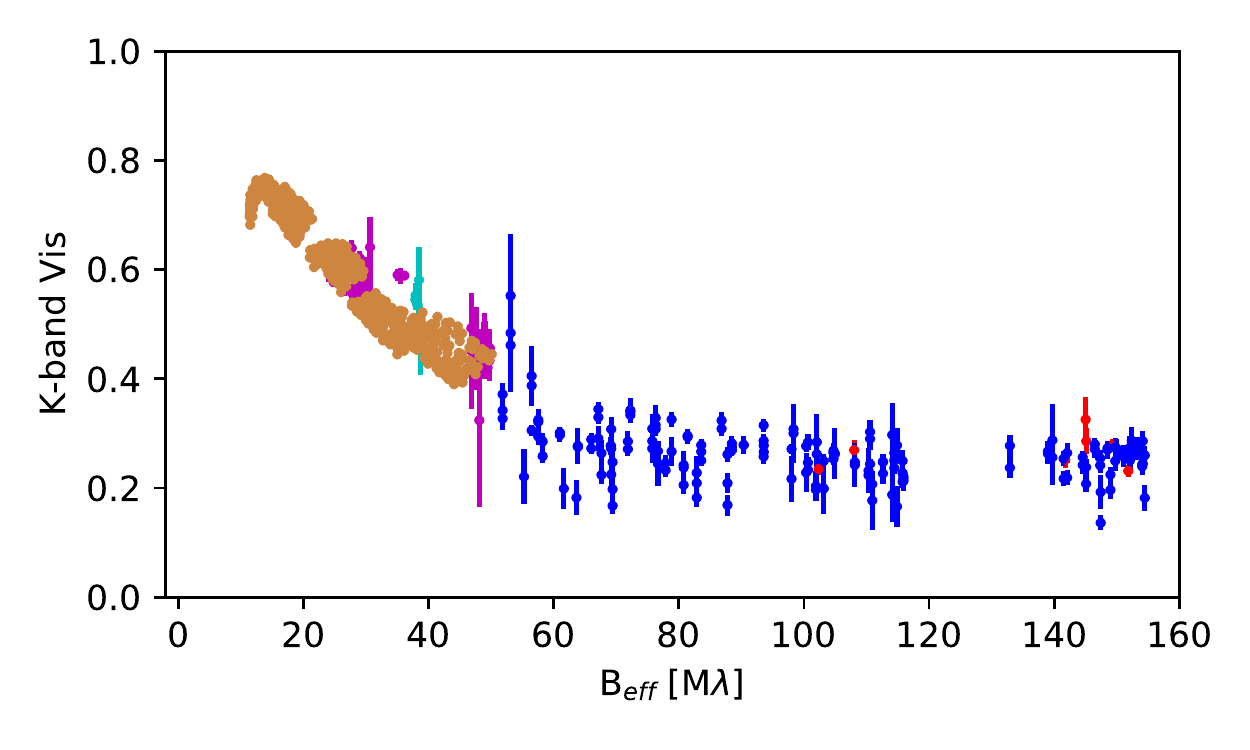}\\
    \includegraphics[trim=0.3cm 0.5cm 0cm 0.3cm, clip=true, width=0.45\textwidth]{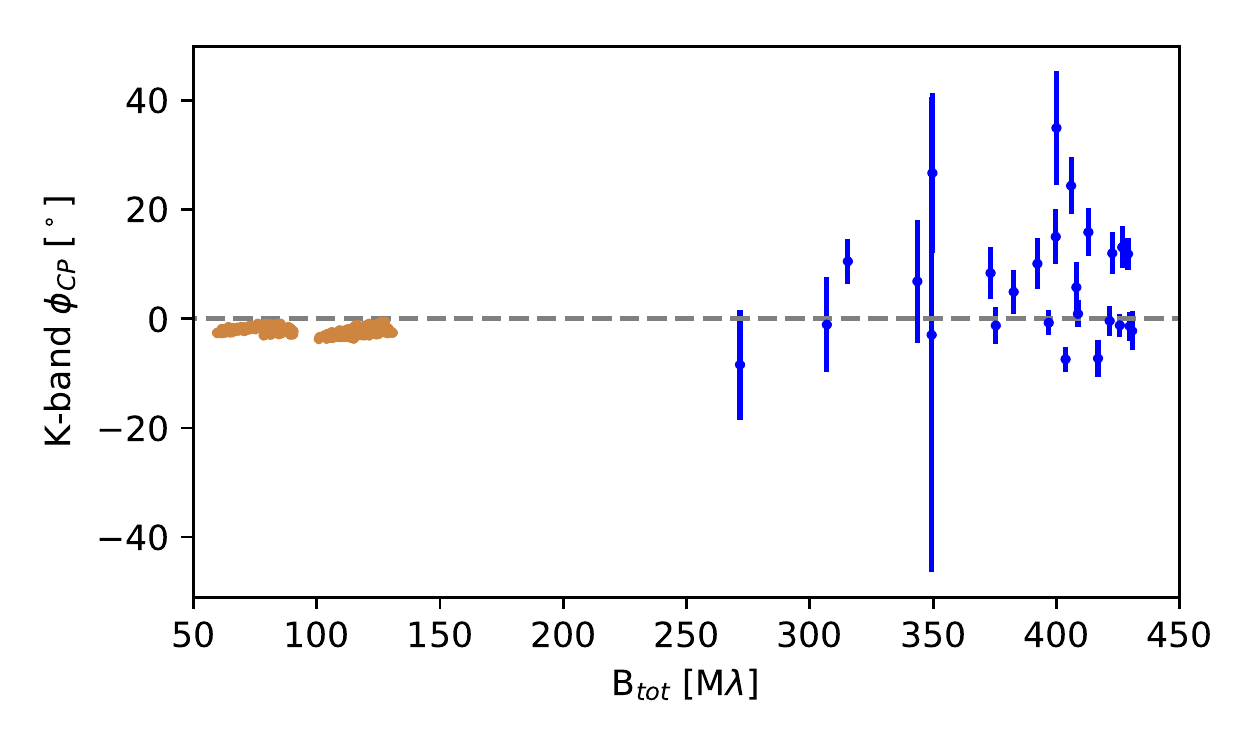}
    \caption{Observed visibilities (top) and $\phi_{\rm{CP}}$ (bottom). Visibilities are plotted with respect to the deprojected baseline length, assuming $i_{\rm{d}}=65^{\circ}$ and PA$_{\rm{major}}=23^{\circ}$ (see Equation~(\ref{eq:beff})). Individual data points are color-coded as in Fig.~\ref{fig:uvplane}.}
    \label{fig:visCP}
\end{figure}

The new and archival visibilities and $\phi_{\rm{CP}}$ obtained for \object{RY~Tau} are displayed in Fig.~\ref{fig:visCP}. Visibilities are plotted with respect to the deprojected baseline length, $B_{\rm{eff}}$, calculated from the baseline vectors using $i_{\rm{d}}=65^{\circ}$, and a disk minor axis position angle, PA$_{\rm{minor}}=113^{\circ}$ (see Section~\ref{sec:results} for details regarding the adopted disk geometry), following
\begin{equation}\label{eq:beff}
    B_{\rm{eff}} = B\left[ \sin^{2}(\phi)+\cos^{2}(i)\cos^{2}(\phi)\right]^{1/2}.
\end{equation}
Here, $\phi$ is the difference between PA$_{b}$ and PA$_{\rm{minor}}$. Using $B_{\rm{eff}}$ rather than the true baseline length, $B$, accounts for the fact that the brightness distribution along PA$_{b}$ which trace PA$_{\rm{minor}}$ is foreshortened in comparison to that along PA$_{b}$ which trace the disk major axis position angle, PA$_{\rm{major}}$.

Before undertaking detailed modeling, we visually inspected the data for signs of temporal variations in the underlying brightness distribution. Specifically, we inspected the vertical spread in visibility with respect to $B_{\rm{eff}}$ (top panel of Fig.~\ref{fig:visCP}). The vertical spread in visibility with $B_{\rm{eff}}$ across the GRAVITY data (orange data points) is dominated by the spectral dependence: the longer wavelength spectral channels display shallower visibility profiles. This effect is consistent with the idea that longer wavelengths probe comparatively cooler regions of the circumstellar disk which are more extended and thus more resolved. In comparison, our CLIMB and CLASSIC data (blue and red data points, respectively) are all obtained using the same filter with no spectral dispersion so spectral variations cannot explain the vertical spread in these data. Splitting the CLIMB data up by observation date does not reveal noticeable temporal variations in the visibility. Instead, a similar level of vertical spread in visibility to that in the top panel of Fig.~\ref{fig:visCP} is present at each observational epoch. We also see no dependence of the CLIMB and CLASSIC visibilities on PA$_{b}$, although we note that our ($u,v$)-plane coverage does not directly probe the $\sim10-15^{\circ}$ region around PA$_{\rm{minor}}$ (see Fig.~\ref{fig:uvplane}). The vertical spread in the CLIMB and CLASSIC data is more likely associated with measurement uncertainty and/or an under-estimation of calibration uncertainties rather than an underlying astrophysical process. Thus, we adopt an additional $10\%$ systematic uncertainty on the CHARA data.

A similar assessment of the potential effect of temporal variability on the $\phi_{\rm{CP}}$ measurements (shown in the bottom panel of Fig~\ref{fig:visCP}) was not possible due to (i) the sparsity of CLIMB data from individual nights and (ii) the availability of only a single epoch of GRAVITY data for comparison. Consequently, we are unable to reliably assess the cause of our non-zero CLIMB $\phi_{\rm{CP}}$ measurements. 

\subsection{Monte Carlo Radiative Transfer models}\label{sec:RTmodel}
We model \object{RY~Tau} as a centrally illuminated passive disk using the TORUS Monte Carlo radiative transfer code \citep{Harries00, Tannirkulam07, Harries19}. In this scenario, viscous heating due to accretion is assumed to be minimal and the central star is the only source of heating. The \citet{Lucy99} algorithm is used to compute radiative equilibrium on a two-dimensional, cylindrical adaptive mesh grid. 

Polarized intensity images of \object{RY~Tau} have previously highlighted a notable scattered light contribution across optical and NIR wavelengths \citep{Takami13, Garufi19}. The distance scales probed by our interferometric observations are much more compact and we anticipated the scattered light contribution to cause the visibilities to deviate from a value of $1.0$ at the shortest baseline lengths. From the top panel of Fig.~\ref{fig:visCP}, it is difficult to assess whether the visibilities are consistent with $1.0$ at zero baseline length. Meanwhile, the GRAVITY visibilities display a ``hook'' feature at the shortest effective baselines which we attribute to our deprojection. Observations probing shorter spatial frequencies are required to assess the flux contribution of any over-resolved component to the visibilities. Here, we assume that this contribution is minimal and that the NIR emission probed by our interferometry arises purely from the sublimation rim at the inner edge of the disk.

We prescribe the density structure of the circumstellar material, $\rho(r,z)$, using the $\alpha$-disk prescription of \citet{Shakura73} whereby
\begin{equation}\label{eq:gasdensity}
 \rho(r, z) = \frac{\Sigma(r)}{h(r)\sqrt{2\pi}} \exp\left\{-\frac{1}{2}\left[\frac{z}{h(r)}\right]^{2}\right\}.
\end{equation}
Here, $r$ and $z$ are the radial distance into the disk and the vertical height above the disk midplane, respectively. The parameters $h(r)$ and $\Sigma(r)$ describe the scale height, 
\begin{equation}\label{eq:h0}
 h(r) = h_{0}\left(\frac{r}{100\,\rm{au}}\right)^{\beta},
\end{equation} 
and the surface density, 
\begin{equation}\label{eq:Sigma}
 \Sigma(r) = \Sigma_{0}\left(\frac{r}{100\,\rm{au}}\right)^{-p},
\end{equation}
of the disk, respectively. Constants $h_{0}$ and $\Sigma_{0}$ are equated at $r=100\,$au. We keep $p=1.0$ fixed in all models. 

In adopting this approach, we do not account for any additional complexity in the radial dependence of the disk surface density and scale height, as evidenced by the apparent dust cavity at $\sim18\,$au seen in CARMA images of RY~Tau \citep{Isella10}, for example. While we do not expect this to affect the modeling of the $K$-band visibilities, we discuss the implications of this approach in relation to the bulk SED in Section~\ref{sec:Sigma_impact}.

The final temperature structure of the disk and the shape of the dust sublimation front are then established in an iterative manner using the \citet{Lucy99} algorithm, provided $T_{\rm{sub}}$ is prescribed for each grain species in the model. We prescribe the disk models using a gas density-dependent sublimation temperature from \citet{Pollack94}:
\begin{equation}
    T_{\rm{sub}} = G\rho^{\gamma}\left(r,z\right).
\end{equation}
Here $G=2000\,$K and $\gamma=1.95\times10^{-2}$. This produces an inner rim that curves away from the star with increasing scale height above/below the disk midplane and whose innermost edge depends on the grains with the largest $T_{\rm{sub}}$ and cooling efficiency \citep{Isella05}. As $T_{\rm{sub}}$ and the cooling efficiency typically increase with increasing grain size, we populate the disk using dust of a single grain size, which we denote $a_{\rm{max}}$, which represents the largest grains which significantly contribute to the opacity in the disk rim. Importantly, this does not mean that grain growth beyond $a_{\rm{max}}$ has not occurred. Instead, any growth of grains beyond $a_{\rm{max}}$ simply does not contribute sufficiently to the opacity in the inner disk. We adopt a single grain model (as in \citealt{Isella05}, for example) as opposed to a two-grain mixture model (as in \citealt{Tannirkulam07}) to control the curvature of the inner rim. This provides a narrower inner disk rim (i.e. one that curves over a smaller range of disk annuli, \citealt{Tannirkulam07}) but which speeds up model computation \citep{Davies18}.

For consistency with \citet{Davies18} and \citet{Labdon19}, only \citet{Draine03} silicates are used. Though this assumption is rather simplistic, it is reasonable considering the good fit provided to the \textit{Spitzer} spectrum by models only considering silicate grains \citep{Espaillat11}. 

\subsubsection{Stellar and bulk disk parameters}\label{sec:starParam}
\begin{deluxetable}{lcccccccc}
\tabletypesize{\scriptsize}
\tablecolumns{8}
\tablecaption{Stellar parameters \label{tab:starParam} }
\tablehead{
\colhead{} & \colhead{$T_{\rm{eff}}$} & \colhead{$\log g$} & \colhead{$d$} & \colhead{$A_{\rm{V}}$} & \colhead{$L_{\star}$} & \colhead{$R_{\star}$} & \colhead{$M_{\star}$} \\
\colhead{} & \colhead{(K)} & \colhead{} & \colhead{(pc)} & \colhead{} & \colhead{($L_{\odot}$)} & \colhead{($R_{\odot}$)} & \colhead{($M_{\odot}$)} }
\startdata
Herein & $5945$ & $3.8$ & $140$ & $1.6$ & $11.6$ & $3.2$ & $2.0$ \\
L19    & $6220$ 
& $4.0$ & $128$ & $1.94$ 
& $12.3$ & 2.37 & $2.04$ 
\\
G19 & $5750$ & $3.58$ & $133$ 
& $1.5$ 
& $6.3$ 
& 3.7 & $\approx1.9$
\enddata
\tablecomments{For ``herein'' row, $T_{\rm{eff}}$, $M_{\star}$, and $\log g$ are from \citet{Calvet04}; $d$ from \citet{Kenyon94, Galli18}; and $A_{\rm{V}}$ from \citet{Petrov19}. See text for details regarding the calculation of $L_{\star}$ and $R_{\star}$. Radii from L19 and G19 have been calculated using $\log g$ and $M_{\star}$.}
\end{deluxetable}

The disk in our TORUS models is passively heated by a single star located at the grid center. Estimates of $T_{\rm{eff}}$, the stellar radius, $R_{\star}$, stellar mass, $M_{\star}$, $d$, and $A_{\rm{V}}$ were required as model inputs. A range of values for \object{RY~Tau}'s stellar parameters have been published and cited throughout the literature - in part due to its photometric and spectroscopic temporal variability. The values adopted herein are presented in Table~\ref{tab:starParam} and a brief discussion of the impact of using commonly adopted alternatives is presented in Section~\ref{sec:star_discussion}. 

$T_{\rm{eff}}$ and $M_{\star}$ are taken from \citet{Calvet04} while we revise their estimate of the stellar luminosity, $L_\star$, using the ``bright'' epoch photometry and $A_{\rm{V}}$ from \citet{Petrov19}. Through analyzing the $V$ versus $(B-V)$ color-magnitude diagram produced using data obtained during their photometric monitoring campaign, \citet{Petrov19} noted that the curved distribution of data points is similar in shape to those of objects exhibiting UX~Ori-type behavior. However, the linear section of data points, which is typically observed for UX~Ori-type objects when the central star is directly observable, is missing. They note that their $A_{\rm{V}}$ estimate -- which is broadly consistent with previous estimates (e.g. \citealt{Calvet04, Herczeg14, Garufi19}) -- likely provides an upper limit for $A_{\rm{V}}$ as a result. From $T_{\rm{eff}}$ and $L_\star$, we re-estimate $R_{\star}$ ($3.2\,\rm{R_\odot}$, see Table~\ref{tab:starParam}). 

As a member of the Taurus star forming region, \object{RY~Tau} is typically considered to be located at $d\sim140\,$pc \citep{Elias78}. In apparent contrast, the estimate of $d$ inferred from the \textit{Gaia} DR2 parallax \citep{Gaia16, Gaia18} suggests a much increased $d=443^{+55}_{-44}\,$pc \citep{Bailer18}. 
However, the renormalized unit weight error (RUWE), provided in \textit{Gaia} DR2 as an assessment of the quality of the astrometric fit for each source \citep{Galli18}, is $6.7$. This indicates a less than ideal astrometric fit, likely related to the strong nebulosity present around \object{RY~Tau} which impacts the \textit{Gaia} point-spread-function. For this reason, we adopt $d=140\,$pc in our modeling of \object{RY~Tau}. 

As our NIR observations (and the SED) are insensitive to the outer disk radius, $R_{\rm{out}}$, we rely on literature estimates of this quantity throughout our modeling, adopting $R_{\rm{out}}=80\,$au \citep{Isella10, Takami13}. Due to the simple grain prescription we adopt, we are also unable to meaningfully estimate the disk mass. Instead, we adopt a total disk mass of $0.3\,\rm{M_{\odot}}$ (assuming a dust-to-gas ratio of 1:100) throughout as this provided a reasonable fit to the sub-mm portion of the SED. 

\subsubsection{Simulated observations}
Following convergence, model SEDs and $K$-band ($\lambda=2.13\mu$m) images were computed using a separate Monte Carlo algorithm based on the optical properties of the specific dust species in each model \citep{Harries19}. Model visibilities were extracted from the images at PA$_{b}=0-180^{\circ}$ and at baseline lengths up to $330\,$m, corresponding to the full range of spatial frequencies probed by our ($u,v$)-plane coverage. The model $\phi_{\rm{CP}}$ were computed from the sum of visibility phases extracted from the image along each closed triangle of baseline vectors (see \citealt{Davies18} for more details). 

\begin{deluxetable}{lcc}
\tabletypesize{\scriptsize}
\tablecolumns{3}
\tablecaption{Prior estimates of the large-scale disk geometry \label{tab:diskParams}}
\tablehead{
& \colhead{$i_{\rm{d}}$} & \colhead{PA$_{\rm{major}}$}\\
& \colhead{($^{\circ}$)} & \colhead{($^{\circ}$)}
}
\startdata
\citet{Agra09}    & $45-76.5$ & $24\pm1$ \\
\citet{Isella10}  & $66\pm2$ & $24\pm3$ \\
\citet{Pinilla18} & 62 & 23 \\
\citet{Long18}    & $65.0\pm0.02$ & $23.06\pm0.02$ \\
L19    & $65.0\pm0.1$  & $23.1\pm0.1$ \\
G19  & $55$ & $23$
\enddata
\tablecomments{Position angles are quoted for the disk major axis and are measured East of North. Parameters from \citet{Agra09} and G19 are determined from the micro-jet orientation and assume the disk plane is perpendicular to this axis. }
\end{deluxetable}

Simulated images and SEDs were computed at $i_{\rm{d}}=65^{\circ}$, based on the estimates of $i_{\rm{d}}$ from mm interferometry (see Table~\ref{tab:diskParams}). Synthetic SEDs were computed at $i_{\rm{d}}=65^{\circ}$ and a near face-on $i_{\rm{d}}=20^{\circ}$, enabling us to asses the level of circumstellar extinction provided by each model. The simulated images were rotated so that PA$_{\rm{major}}=23^{\circ}$ East of North and the brighter side of the disk in each simulated image lay to the North West to match the images obtained with HST \citep{Agra09}, VLT/SPHERE (G19) and Subaru/HiCIAO \citep{Takami13}. PA$_{\rm{minor}}$ ($113^{\circ}$) is also in good agreement with the micro-jet axis position angle observed by HST \citep{StOnge08, Agra09} and VLT/SPHERE (G19) indicating no strong evidence for misalignment between the inner and outer disk regions. 

\subsection{The nature of the circumstellar $K$-band emission}\label{sec:results}
\begin{figure*}
    \centering
    \includegraphics[trim=2.5cm 16.5cm 2.5cm 2cm, clip=true,width=0.95\textwidth]{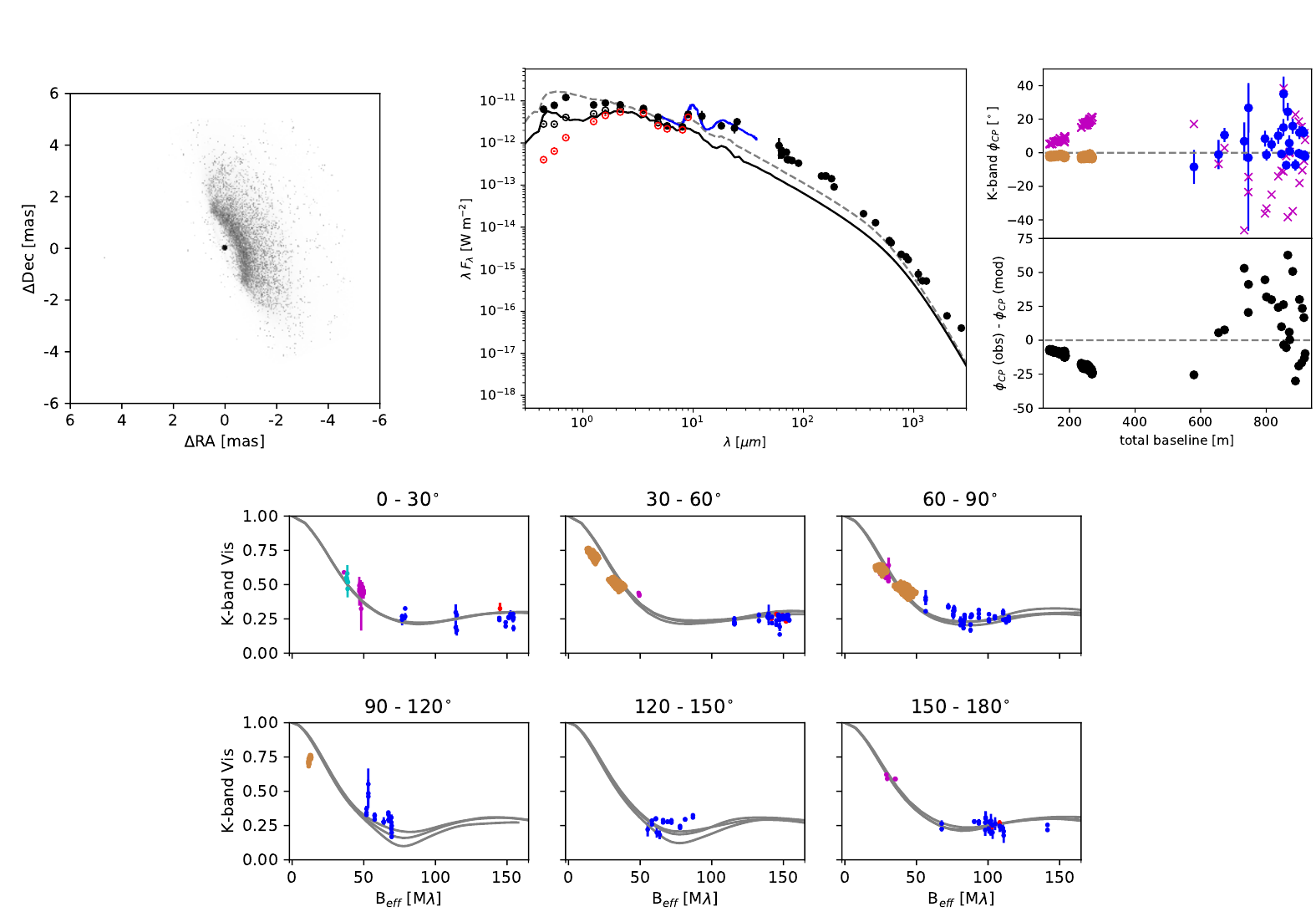}
    \caption{TORUS model providing the best fit to the visibilities ($h_{\rm{0}}=8\,$au; $\beta=0.99$ and $a_{\rm{max}}=0.40\,\mu$m). The TORUS model $2.13\,\mu$m image (top left) was computed at $i_{\rm{d}}=65^{\circ}$ and rotated such that PA$_{\rm{major}}=23^{\circ}$. The SED (top middle) compares the data from Fig.~\ref{fig:sed_stellar_direct020} to the TORUS model computed at $i_{\rm{d}}=65^{\circ}$ (solid black line) and a more face-on $i_{\rm{d}}=20^{\circ}$ (dashed grey line) to highlight the amount of local extinction provided by the disk rim. The $\phi_{\rm{CP}}$ (upper panel) and their residuals (lower panel) are displayed in the top right. Orange and blue data points have the same meaning as in Fig.~\ref{fig:uvplane} while pink crosses indicate the model values extracted from the image. The lower two panels show the visibilities (colors as in Fig.~\ref{fig:uvplane}) compared to the visibility curves extracted from the model image at increments of $10^{\circ}$ in PA$_{b}$ (solid grey lines). Visibilities are split according to PA$_{b}$ (the range is labeled above each subplot).}
    \label{fig:Bestmod}
\end{figure*}

The stellar (Table~\ref{tab:starParam}) and bulk disk parameters ($R_{\rm{out}}=80\,$au; $M_{\rm{disk}}=0.30\,\rm{M_{\odot}}$; see Section~\ref{sec:starParam}) were kept fixed throughout our TORUS modeling. We investigated different values of the maximum grain size contributing to the opacity in the inner disk, $a_{\rm{max}}$, together with the scale height constant, $h_{\rm{0}}$, and flaring parameter, $\beta$. Together, these variables control the location, size, and shape of the NIR-emitting inner disk. 

We performed an initial exploration of a broad range of model parameters to explore their interdependence. We assessed the goodness-of-fit of each model using the following procedure\footnote{We note that the SED beyond NIR wavelengths was largely ignored in this procedure as we do not expect equations~(\ref{eq:h0}) and (\ref{eq:Sigma}) to fully prescribe the radial dependence of the scale height and the surface density, respectively.}: 
\begin{enumerate}
    \item the model visibilities were inspected by-eye to check for consistency with the overall shape of the observed visibilities and the minimum observed visibility level;
    \item the model SED across optical and NIR wavelengths was compared to the data to ensure it fell within the range between the ``bright'' and ``faint'' epoch optical and NIR photometry;
    \item if the model passed these checks, the goodness-of-fit of the model to the visibilities was evaluated using the $\chi_{\rm{r}}^{2}$ statistic.
\end{enumerate} 
These assessments were then used to select the values to be explored on the next iteration of models. This resulted in a sparsely sampled set of models with $a_{\rm{max}}$ ranging between $0.10$ and $1.20\,\mu$m, $h_{\rm{0}}$ ranging between $4$ and $14\,$au, and $\beta$ ranging between $0.88$ and $1.40$. In total, we explored $\sim150$ different combinations of values for these parameters. 

\begin{figure}
    \centering
    \includegraphics[trim=0cm 0cm 0cm 0.0cm, clip=true,width=0.4\textwidth]{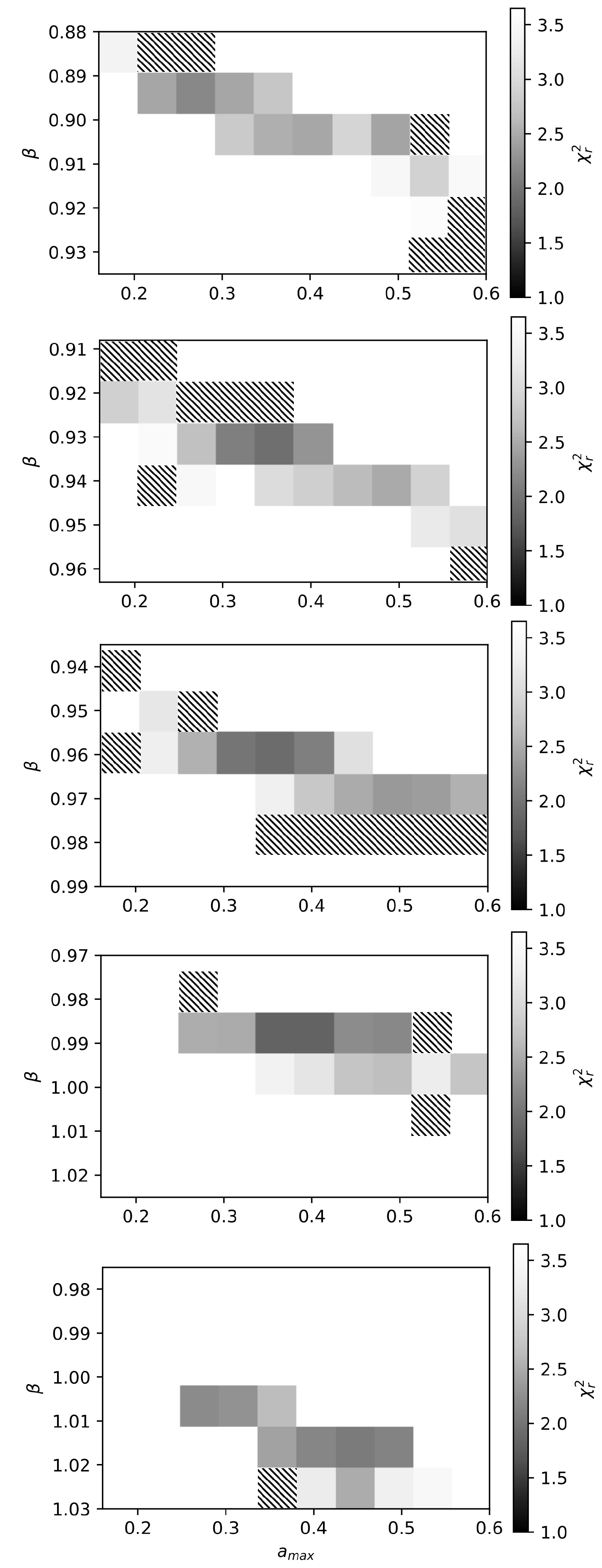}
    \caption{$\chi^{2}_{\rm{r}}$ maps for models with different maximum grain sizes ($a_{\rm{max}}$; x-axis) and scale height constants (from top to bottom: $h_{0}=5\,$au, $6\,$au, $7\,$au, $8\,$au, and $9\,$au) and flaring parameters ($\beta$; y-axis) when considering fits to all the data (i.e. $798$ degrees of freedom) and a 10\% systematic error on the visibility measurements. Models which provided poor fits to the data (i.e. $\chi^{2}_{\rm{r}}$ exceeded the range plotted) are shown as hatched boxes.}
    \label{fig:chisq_map_all}
\end{figure}

\begin{figure}
    \centering
    \includegraphics[trim=0cm 0cm 0cm 0.0cm, clip=true,width=0.4\textwidth]{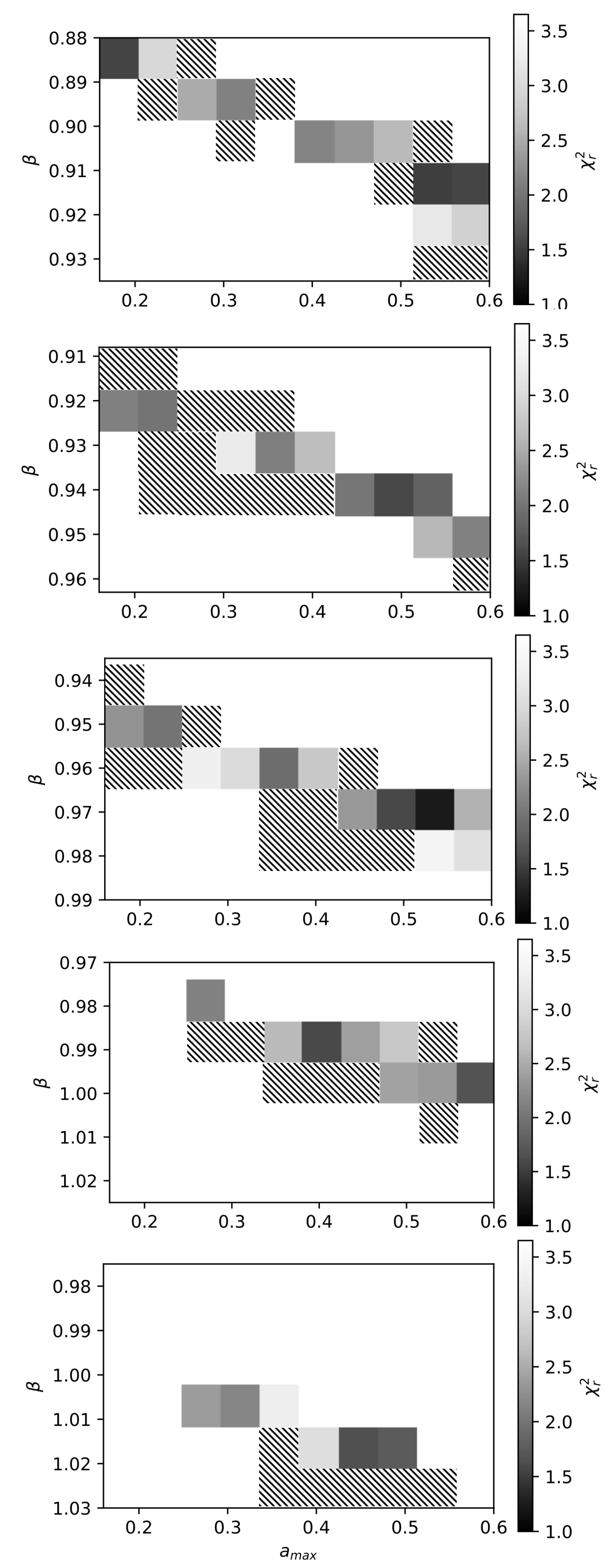}
    \caption{As Fig.~\ref{fig:chisq_map_all} but considering only the CHARA (CLASSIC \& CLIMB) data in the fitting process (i.e. $171$ degrees of freedom). The additional 10\% systematic error to the visibilities is still included.}
    \label{fig:chisq_map_chara}
\end{figure}

\begin{figure*}
    \centering
    \includegraphics[trim=2.5cm 16.5cm 2.5cm 2cm, clip=true,width=0.95\textwidth]{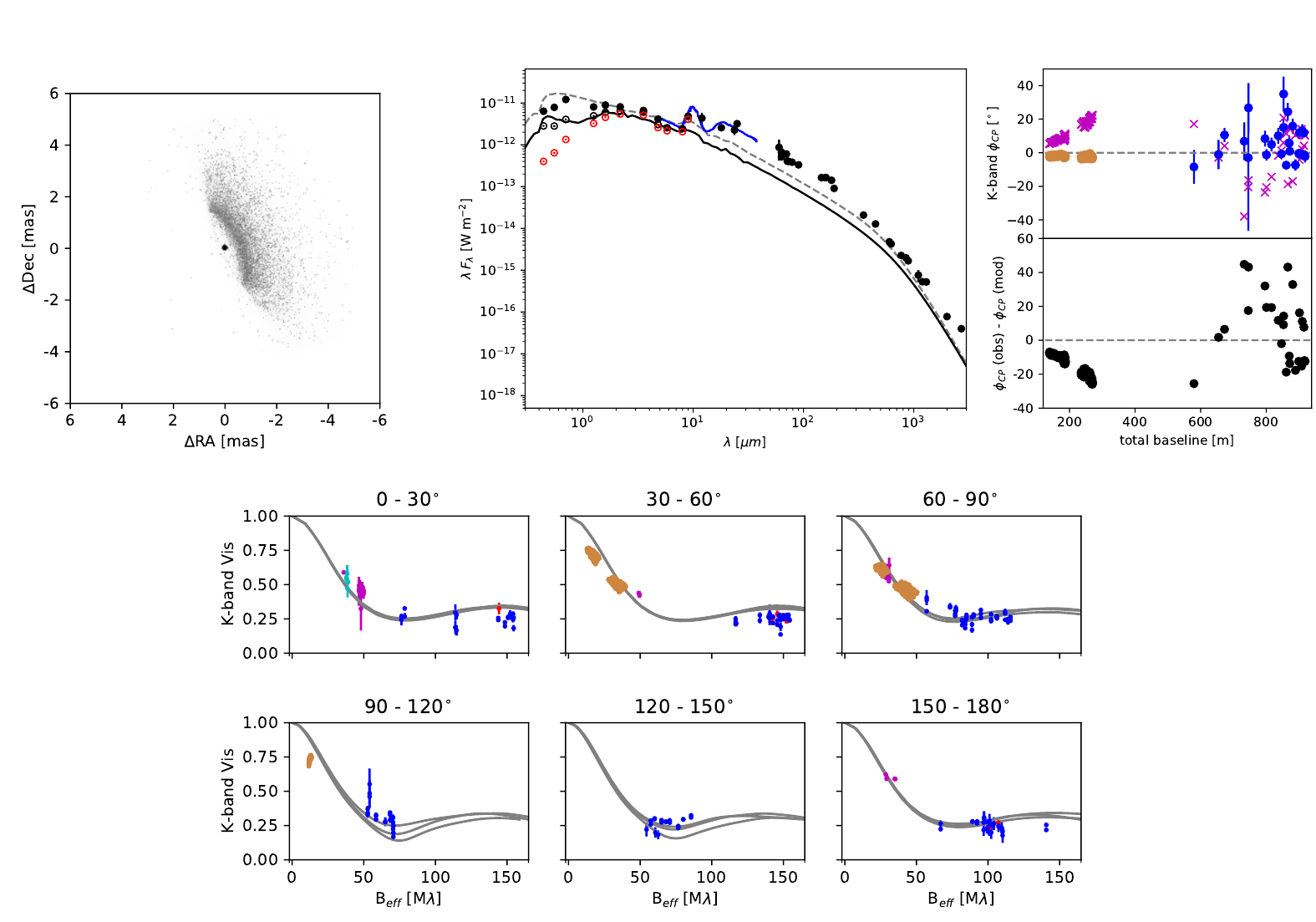}
    \caption{As Fig.~\ref{fig:Bestmod} but for the TORUS model with $h_{\rm{0}}=8\,$au; $\beta=0.99$ and $a_{\rm{max}}=0.36\,\mu$m).}
    \label{fig:2ndBestMod}
\end{figure*}

Based on the results from this initial suite of models, we refined our model exploration around promising regions of the $a_{\rm{max}}$--$h_{\rm{0}}$--$\beta$ parameter space and computed a finer grid of TORUS models with $5\leq h_{\rm{0}}\leq 9\,$au, $0.88\leq \beta \leq 1.03$, and $0.16\leq a_{\rm{max}}\leq0.60\,\mu$m. Above $h_{\rm{0}}\approx 9\,$au, we found models were unable to simultaneously fit the visibilities and the optical-to-NIR region of the SED. Instead, the NIR flux in the SED was consistently underestimated by the models, even when the model visibilities provided a reasonable fit to those observed. Meanwhile, a surprising behavior of models with low $a_{\rm{max}}$ set the lower limit to the range of $h_{\rm{0}}$ values we explored: for models with $a_{\rm{max}}\lesssim0.16\,\mu$m we found the inner edge of the disk rim moves \emph{inward} with decreasing $a_{\rm{max}}$ (see Appendix~\ref{apen:grains}), opposite to what happens for larger grains ($a_{\rm{max}}\gtrsim0.16\,\mu$m). This effect has not been reported by studies conducting similar analyses for hotter stars (e.g. \citealt{Isella05, Davies18}). Further investigation of this effect is outside the scope of this paper and is deferred to future study (Davies \& Harries 2020, in preparation). 

Our analysis shows that the circumstellar $K$-band emitting region is consistent with models of a disk inner edge shaped by dust sublimation. Our best-fit model suggests that the dust rim extends inwards to within $0.206\pm0.001\,$au of the central star, corresponding to $R_{\rm{sub}}$ for Silicate dust with $a_{\rm{max}}=0.40\,\mu$m. Specifically, the model providing the best fit to the visibility data has $a_{\rm{max}}=0.40\,\mu$m, $h_{\rm{0}}=8\,$au, and $\beta=0.99$. The quoted uncertainty of $\pm0.001\,$au is equivalent to half a grid cell on our adaptive mesh (see Section~\ref{sec:rsub_error} for a more detailed discussion of the uncertainty on our assessment of $R_{\rm{sub}}$). 

We display the corresponding TORUS model image (top left panel), SED (top middle panel), $\phi_{\rm{CP}}$ (top right panel) and visibilities (lower panels) in Fig.~\ref{fig:Bestmod}. The visibilities are split by PA$_{b}$ to show the relative goodness-of-fit across different segments of the disk. The full SED (from optical to millimeter wavelengths) is shown to illustrate the goodness-of-fit of our model across the optical and NIR while also demonstrating how our adoption of radial power laws for the scale height and surface density (equations~\ref{eq:h0}) and (\ref{eq:Sigma}), respectively) underestimates the flux at longer wavelengths. We discuss the implications of this in more detail below.

\section{Discussion}\label{sec:discussion}
\subsection{RY~Tau as a pre-transitional disk}\label{sec:Sigma_impact}
Figure~\ref{fig:Bestmod} clearly shows that our best-fit model provides a poor fit to the SED at wavelengths longer than $\sim10\,\mu$m. This behavior is seen across all of the models we explored and is not unexpected: the shape of the SED of \object{RY~Tau} has led previous studies to classify the circumstellar structure as a pre-transitional disk \citep{Furlan09, Espaillat11}. Moreover, CARMA and ALMA images of \object{RY~Tau} have highlighted the likely presence of a dust cavity or gap within $\sim18$au (\citealt{Isella10, Long18}; L19). The disk structure is thus expected to deviate from the radial power laws we have used for the scale height and the surface density in our TORUS models (equations~\ref{eq:h0} and \ref{eq:Sigma}, respectively). In addition, extrapolating the single grain size dust model of the inner dust rim to the full disk impacts on the outer disk emissivity as well as the strength and shape of the silicate feature. 

\subsection{Uncertainty estimate for $R_{\rm{sub}}$}\label{sec:rsub_error}
We display the comparative goodness-of-fit of our grid of models to the observed visibilities as $\chi^{2}_{\rm{r}}$ maps in Fig.~\ref{fig:chisq_map_all}. Hatched grid cells highlight areas of the map that provided poorer fits to the data than the maximum $\chi^{2}_{\rm{r}}$ value indicated by the colorbar. White cells highlight unexplored regions of our model parameter space. As the number of data points provided by the GRAVITY observations far exceeds the number provided by CLASSIC and CLIMB, the short baseline data dominate the assessment of the goodness-of-fit. To combat this, we also calculated the goodness-of-fit to the CHARA visibilities only (Fig.~\ref{fig:chisq_map_chara}). 

Some models which used different combinations of $h_{\rm{0}}$, $\beta$ and/or $a_{\rm{max}}$ produced similarly good fits to the visibilities. In these models, the steeper increase in scale height with respect to disk radius, provided by decreasing $\beta$, was counteracted by the reduction in $h_{\rm{0}}$ (or vice versa). For example, the model with $h_{\rm{0}}=8\,$au, $\beta=0.99$, and $a_{\rm{max}}=0.36\,\mu$m (see Fig.~\ref{fig:2ndBestMod}) produces only a marginally poorer fit to the short baseline visibilities ($\chi^{2}_{\rm{r}}=1.843$ compared to $\chi^{2}_{\rm{r}}=1.841$) while the goodness-of-fit provided to the CHARA visibilities is poorer than in our best-fit model ($\chi^{2}_{\rm{r}}=2.633$ compared to $\chi^{2}_{\rm{r}}=1.606$). 

The short baseline data are the most sensitive to $a_{\rm{max}}$ as they trace the fall-off in visibility with increasing spatial frequency. From Fig.~\ref{fig:chisq_map_all}, we see that our assessment of $a_{\rm{max}}$ is reasonably robust. With the exception of the $h_{0}=5\,$au models, which all provide similarly poor fits to the visibilities, the best-fitting model in each $\chi^{2}_{\rm{r}}$ map has $a_{\rm{max}}=0.36-0.40\,\mu$m. Based on the results from all of our models, we estimate $R_{\rm{sub}}=0.210\pm0.005\,$au. Larger grains produce inner disk rims that are under-resolved compared to our data while smaller grains produce comparatively over-resolved rims. However, it is important to note that this result does not rule out grain growth to larger sizes. Instead, our result indicates that the number density of Silicate grains larger than $0.40\,\mu$m in the inner disk rim is insufficient for these grains to contribute significantly to the opacity at the inner disk. Furthermore, due to their associated optical properties, our observations are insensitive to the presence of grains larger than $\sim1.2\,\mu$m \citep[c.f.][]{Isella05}.

\subsection{Sensitivity of our results to the adopted stellar input parameters}\label{sec:star_discussion}
Our modeling in previous sections relies to a certain extent on the assumption that the stellar parameters we have adopted are representative of the true values. As we outlined in Section~\ref{sec:starParam}, accurately assessing values for $T_{\rm{eff}}$, $L_{\star}$, $d$, and $A_{\rm{V}}$ for \object{RY~Tau} is complicated by photometric variability and direct occultation by the disk surface layers. In this subsection, we briefly assess the sensitivity of our results to the stellar parameters adopted. 

In Table~\ref{tab:starParam}, we provide example alternative stellar parameters for \object{RY~Tau}, recently adopted in L19 and G19. L19 co-added 96 archival ESPaDOnS spectra and compared them to F and G spectral type BT-Settl models with solar metallicity and surface gravity, $\log g=4.0$. They yielded $T_{\rm{eff}}=6220\pm80\,$K (comparable to F6-F8 spectral types using \citet{Kenyon95} spectral type-to-$T_{\rm{eff}}$ relations). This is a small change in spectral type from the more commonly adopted values of G0 \citep{Herczeg14} and G1 \citep{Calvet04}. G19 also re-estimated $T_{\rm{eff}}$, comparing four archival high resolution William Herschel Telescope UES spectra to synthetic atmosphere models computed from the ATLAS and SYNTHE codes and finding $T_{\rm{eff}}=5750\,$K with $\log g=3.58$, closer to our adopted values ($T_{\rm{eff}}=5945\,$K with $\log g=3.8$). 

L19 estimated $A_{\rm{V}}=1.94\pm0.2\,$mag, higher than our adopted value of $1.6\,$mag while G19 estimated $A_{\rm{V}}=1.5$mag. For $d$, L19 and G19 both assessed the \textit{Gaia} parallaxes of the 29 closest Taurus members to \object{RY~Tau}, computing an average \textit{Gaia} distance of $128.5\pm0.3\,$pc. L19 adopted this value for \object{RY~Tau} while G19 used this calculation to argue the case for adopting the Hipparcos value ($d=133\,$pc). Based on these differences, the estimates of $L_{\star}$ from these two studies then differ greatly with L19 estimating $L_{\star}=12.3\,\rm{L_{\odot}}$ and G19 estimating $L_{\star}=6.3\,\rm{L_{\odot}}$. 

Combined with our best-fitting disk model from Section~\ref{sec:results}, the different stellar input parameters produce model SEDs with similar shapes but different intensities. The model using G19 stellar parameters has an $R_{\rm{sub}}$ consistent with our estimate above ($0.212\,$au). Meanwhile, the flux across the IR provided by the L19 model underestimates that in the SED compiled from archival photometry. If these stellar parameters are closer to \object{RY~Tau}'s true values, this indicates that less of the line-of-sight extinction is provided by circumstellar material than in our best-fit disk model. Using our best-fit disk model with L19 stellar parameters produces a less-extended inner rim, with $R_{\rm{sub}}\approx0.166\,$au. The poor fit to the visibilities provided by this model indicates that this is not a good estimate. Decreasing $a_{\rm{max}}$ to $0.20\,\mu$m provides an improved fit with $R_{\rm{sub}}\approx0.210$ once again. Thus, it appears our estimate of $R_{\rm{sub}}$ is reasonably robust against differences in stellar parameter estimates.

\subsection{Comparison of $R_{\rm{sub}}$ to the theoretical magnetospheric truncation radius}
To further characterize the inner disk of \object{RY~Tau}, we calculate and compare the magnetospheric truncation radius, $R_{\rm{trunc}}$, to the value of $R_{\rm{sub}}$ inferred from our TORUS modeling. The magnetospheric truncation radii of Herbig Ae stars are typically far interior to $R_{\rm{sub}}$, leaving a portion of the inner disk completely devoid of Silicate grains\footnote{We explicitly mention Silicate grains here as, if more refractory grains are present, they will be able to survive closer to the star at higher temperatures.}. However, for lower mass, T-Tauri stars, the locations of $R_{\rm{trunc}}$ and $R_{\rm{sub}}$ may overlap, leading to the possibility of dust being lifted into magnetospheric accretion streams (e.g. \citealt{Bodman17}) and producing a warped inner disk where the scale height, measured with respect to a reference disk midplane, varies with azimuth (e.g. \citealt{Kesseli16}). In light of this, we calculate $R_{\rm{trunc}}$ and compare it our estimate of $R_{\rm{sub}}$ to assess the applicability of the azimuthally invariant scale height prescription (equation~(\ref{eq:h0})). 

Considering the force balance between the outward pressure from the large-scale stellar magnetic field, $B_{\star}$, and the inward pressure from mass accretion through the disk (e.g. \citealt{Johnstone14}),
\begin{equation}\label{eq:rtrunc}
R_{\rm{trunc}} = c(2GM_{\star})^{-1/7}\dot{M}_{\rm{acc}}^{-2/7}\mu_{1}^{4/7}.
\end{equation}
Here, $G$ is the gravitational constant, $\dot{M}_{\rm{acc}}$ is the mass accretion rate through the disk and $\mu_{1}$ is the dipole moment\footnote{This equation implicitly assumes the adoption of cgs units.}. The constant, $c$, accounts for the difference between spherical infall and magnetospheric accretion along columns. If $B_{\star}$ is dominated by dipolar fields (a good approximation at sufficient distances from the star due to the increased fall-off with radius of higher order fields) and the disk axis is perpendicular to the stellar magnetic field axis, $c=0.5$ \citep{Long05} and $\mu_{1}=B_{\rm{dip}}R_{\star}^{3}$ at equatorial regions. Here, $B_{\rm{dip}}$ is the strength of the dipole component of $B_{\star}$ at the stellar equator and $R_{\star}$ the stellar radius, as before. We note that in reality, higher order fields become important for small $R_{\rm{trunc}}$ (i.e. high mass accretion rates or low magnetic field strengths, for a given $M_{\star}$; \citealt{Gregory16}) but we only consider the case of a dipole field here for simplicity. 

\object{RY~Tau} was observed using Zeeman-Doppler imaging as part of the Magnetic Protostars and Planets (MaPP) project (PI: J.-F. Donati) with a dipole magnetic field strength, $B_{\rm{dip}}\sim300\,$G measured in preliminary analysis (J.-F. Donati, private communication). Assuming the stellar mass accretion rate ($6.4-9.1\times10^{-8}\,\rm{M_{\odot}yr^{-1}}$; \citealt{Calvet04}) is a good first approximation for $\dot{M}_{\rm{acc}}$, we estimate $R_{\rm{trunc}}\approx0.009-0.014\,$au. This is an order of magnitude closer to the star than our estimate of $R_{\rm{sub}}$ ($0.210\pm0.005\,$au), indicating we are fine to assume an azimuthally symmetric scale height prescription to the inner disk edge.

\section{Conclusions}\label{sec:conclusion}
We find that the $K$-band visibilities and optical-to-NIR SED of \object{RY~Tau} are consistent with Monte Carlo radiative transfer models comprising a central star illuminating a passive disk with an inner edge shaped by dust sublimation with $R_{\rm{sub}}=0.210\pm0.005\,$au. The location of the inner rim is consistent with the sublimation radius of a disk where the largest grains contributing to the opacity (and thus controlling the rim location) are $0.36-0.40\,\mu$m. The growth of dust grains beyond $0.40\,\mu$m cannot be ruled out but our results show that such grains do not contribute significantly to the opacity in the inner rim of the disk.

Interestingly, \citet{Labdon19} found that the location of the inner disk of SU~Aur is similarly controlled by the sublimation of $0.40\,\mu$m grains while \citet{Davies18} found that larger ($1.2\,\mu$m) grains were required to reproduce their $H$- and $K$-band interferometric observations of HD~142666. Both SU~Aur and HD~142666 are similar in mass ($\sim2\,\rm{M_{\odot}}$) to \object{RY~Tau} while HD~142666 is older ($>10\,$Myr; \citealt{Dionatos19}) and more luminous ($\sim20\,\rm{L_{\odot}}$; \citealt{Davies18}) than SU~Aur and \object{RY~Tau} (both $\sim2\,$Myr as members of the Taurus-Auriga star forming region \citep{Luhman18} and $\sim12\,\rm{L_{\odot}}$). Similar analyses of a greater number of disk-hosting YSOs is required before we can comment on whether this is possibly symptomatic of, for example, an evolutionary sequence for disks or that dust grains have to be larger to have survived as long as they have done around HD~142666.

While our models provide a good fit to the optical-to-NIR portion of the SED of \object{RY~Tau}, they consistently poorly fit the data at longer wavelengths ($\gtrsim10\,\mu$m). This is due to the combined effect of populating our disk models with dust of a single grain size and assuming the disk temperature and density can be prescribed using simple radial power laws (Section~\ref{sec:RTmodel}). Previous analysis of the SED \citep[e.g.][]{Furlan09, Espaillat11} and mm interferometry \citep{Isella10} of \object{RY~Tau} has revealed the presence of at least one annular cavity at a separation of $\sim18\,$au from the central star. Thus, there is likely a deviation from simple radial power laws in temperature and density at a certain disk radius. MIR interferometric observations of \object{RY~Tau} with the VLTI's MATISSE instrument \citep{Lopez14}, for example, are required to further assess the structure of the disk between the sublimation rim and the outer disk regions probed by CARMA and ALMA. 

We used existing measurements of the mass accretion rate and large-scale dipolar magnetic field strength of \object{RY~Tau} to estimate a disk truncation radius of $0.009-0.014\,$au. This indicates that, while the dusty portion of the disk has an inner boundary at $0.210\pm0.005\,$au due to sublimation, the gaseous portion of the disk may theoretically extend an order of magnitude closer to the star. Furthermore, this also validates our assumption of an azimuthally symmetric dust rim as it shows that dust is unlikely to survive close enough to the star to be lofted into magnetospheric accretion streams. 

Our CHARA data was obtained over a four year period but our analysis reveals no direct evidence of temporal variability in the disk of \object{RY~Tau}. Instead, the vertical spread in visibility across baselines probed by our CHARA observations is more likely attributed to measurement and calibration uncertainties. However, our exploration of the $a_{\rm{max}}$--$h_{\rm{0}}$--$\beta$ model parameter space in Section~\ref{sec:results} highlights that models which produce a disk that is too shallow or too extended to directly occult the central star can be ruled out. These models consistently overestimate the visibilities on the baselines probed by our CHARA observations, indicating the stellar contribution to the flux contrast in the underlying brightness distribution is too high. In their analysis of \object{RY~Tau}'s photometric variability, \citet{Petrov19} drew similar conclusions and suggested the observer's line of sight to the stellar photosphere was partially occulted even during \object{RY~Tau}'s brightest epochs. Furthermore, our results support previous claims based on (i) the timescales of quasi-periodic optical brightness variations \citep{Zajtseva10}; (ii) the correlation between outflow velocity and circumstellar accretion \citep{Babina16} and (iii) seesaw-like variability in the \textit{Spitzer} spectrum \citep{Espaillat11} that it is the surface layers of the inner disk, close to the dust sublimation rim, that provides this occulting surface.

While we are unable to comment on the possible intrinsic variability of the central star, the direct line-of-sight occultation of the star by the disk provides a mechanism by which structural changes in the surface layers of the dusty portion of the disk can give rise to the aperiodic brightness fluctuations observed across optical and IR wavelengths. The increased sensitivity of the six-telescope MIRC-X combiner \citep{Kraus18,Anugu18} at the CHARA Array provides an exciting opportunity to search for such structural changes in the disk of this object and others showing aperiodic photometric variability.

\acknowledgments
We thank the anonymous referee whose comments ensured greater clarity in the presentation of our results. CLD and SK acknowledge support from the ERC Starting Grant ``ImagePlanetFormDiscs'' (Grant Agreement No. 639889). R.G.L. acknowledges support by Science Foundation Ireland under Grant No. 18/SIRG/5597. We thank Bernard Lazareff, Jean-Baptiste Le Bouquin and Rachel Akeson for their assistance in acquiring archival data and University of Exeter summer research project student Daniel J. Barker for his assistance in incorporating automatic VisIt plotting functionality into the analysis pipeline. CLD thanks Aaron Labdon, Scott Gregory, Jean-Francois Donati, Francois Menard and Catherine Dougados for helpful discussions.
This work is based upon observations obtained with the Georgia State University Center for High Angular Resolution Astronomy (CHARA) Array at Mount Wilson Observatory and data obtained from the ESO Science Archive Facility. The CHARA Array is supported by the National Science Foundation under Grant No. AST-1636624 and AST-1715788. Institutional support has been provided from the GSU College of Arts and Sciences and the GSU Office of the Vice President for Research and Economic Development.
The calculations for this paper were performed on the University of Exeter Supercomputer, a DiRAC Facility jointly funded by STFC, the Large Facilities Capital Fund of BIS, and the University of Exeter.
This research has made use of: the NASA/IPAC Infrared Science Archive, which is operated by the Jet Propulsion Laboratory, California Institute of Technology, under contract with the National Aeronautics and Space Administration (NASA); the Keck Observatory Archive (KOA), which is operated by the W. M. Keck Observatory and the NASA Exoplanet Science Institute (NExScI), under contract with NASA; the Jean-Marie Mariotti Center \texttt{OiDB} service\footnote{Available at http://oidb.jmmc.fr }; the SIMBAD database, operated at CDS, Strasbourg, France; the VizieR catalogue access tool, CDS, Strasbourg, France; NASA's Astrophysics Data System Bibliographic Services. This work has made use of services produced by the NASA Exoplanet Science Institute at the California Institute of Technology.
The Palomar Testbed Interferometer was operated by the NASA Exoplanet Science Institute and the PTI collaboration. It was developed by the Jet Propulsion Laboratory, California Institute of Technology with funding provided from NASA.

%

\facilities{VLTI, CHARA, Keck, PTI}.

\software{TORUS \citep{Harries19}, 
          pysynphot \citep{pysynphot}, 
          NumPy \citep{van2011numpy}, 
          matplotlib \citep{Hunter07}, 
          Astropy \citep{astropy}
          }




\appendix

\section{Multi-band photometry used to build the SED}\label{apen:phot}
The multi-band photometry used to build the SED of \object{RY~Tau}, together with their individual references, are shown in Tables~\ref{tab:phot_bf} and \ref{tab:phot:RY}.

\begin{deluxetable}{ccccccccc}
\tabletypesize{\scriptsize}
\tablecolumns{9}
\tablecaption{Adopted ``bright'' and ``faint'' optical and IR photometric magnitudes, taken from \citet{Petrov19}. \label{tab:phot_bf}}
\tablehead{
\colhead{Date} & \colhead{B} & \colhead{V} & \colhead{R} & \colhead{J} & \colhead{H} & \colhead{K} & \colhead{L} & \colhead{M}}
\startdata
1989 Oct 25 & 11.20 & 10.09 &  8.96 & 7.15 & 6.12 & 5.26 & 4.09 & 3.70\\
2016 Nov 11 & 12.08 & 11.21 & 10.15 & 7.68 & 6.55 & 5.50 & 4.19 & 4.03
\enddata
\end{deluxetable}

\begin{deluxetable}{ccl}
\tabletypesize{\scriptsize}
\tablecolumns{3}
\tablewidth{0pc}
\tablecaption{Additional photometry retrieved from the literature with measurement uncertainties where reported. \label{tab:phot:RY} }
\tablehead{
\colhead{$\lambda$} & \colhead{Flux}   & \colhead{Reference} \\
\colhead{($\mu$m)}     & \colhead{(Jy)}     & \colhead{} }
\startdata
$5.8$ & $4.2$ & \citet{Cieza09}\\
$8.0$ & $5.50$ & \citet{Cieza09}\\
$9.0$ & $12.28\pm0.07$ & \citet{Abrahamyan15}\\
$12.0$ & $12.73$ & \citet{Moshir90}\\
$18.0$ & $15.43\pm0.14$ & \citet{Abrahamyan15}\\
$23.68$ & $17.86\pm4.42$ & \citet{Robitaille07}\\
$25.0$ & $26.70\pm5.00$ & \citet{Moshir90}\\
$60.0$ & $17.40\pm9.00$ & \citet{Moshir90}\\
$63.0$ & $14.10\pm0.05$ & \citet{Keane14}\\
$63.18$ & $10.86\pm0.07$ & \citet{Howard13}\\
$70.0$ & $14.13\pm1.40$ & \citet{Howard13}\\
$71.42$ & $9.63\pm0.96$ & \citet{Robitaille07}\\
$72.84$ & $9.82\pm0.03$ & \citet{Howard13}\\
$78.74$ & $10.10\pm0.04$ & \citet{Howard13}\\
$90.16$ & $10.00\pm0.04$ & \citet{Howard13}\\
$100.0$ & $36.50\pm25.00$ & \citet{Moshir90}\\
$145.53$ & $7.98\pm0.02$ & \citet{Howard13}\\
$157.74$ & $8.64\pm0.03$ & \citet{Howard13}\\
$160.0$ & $8.81\pm0.88$ & \citet{Howard13}\\
$179.53$ & $8.50\pm0.04$ & \citet{Howard13}\\
$189.57$ & $5.73\pm0.11$ & \citet{Howard13}\\
$350.0$ & $2.44\pm0.33$ & \citet{Andrews05}\\
$450.0$ & $1.92\pm0.16$ & \citet{vanderMarel16}\\
$600.0$ & $0.96\pm0.04$ & \citet{Mannings94}\\
$624.0$ & $0.89\pm0.14$ & \citet{Beckwith91}\\
$769.0$ & $0.58\pm0.04$ & \citet{Beckwith91}\\
$850.0$ & $0.56\pm0.03$ & \citet{vanderMarel16}\\
$890.0$ & $0.50\pm0.03$ & \citet{Andrews13}\\
$1100.0$ & $0.28\pm0.09$ & \citet{Mannings94}\\
$1200.0$ & $0.21\pm0.02$ & \citet{Altenhoff94}\\
$1300.0$ & $0.227\pm0.007$ & \citet{Isella10}\\
$2000.0$ & $0.052\pm0.006$ & \citet{Kitamura02}\\
$2700.0$ & $0.036\pm0.003$ & \citet{Isella10}
\enddata
\end{deluxetable}

\section{Sublimation rim location dependence on grain size}\label{apen:grains}
\begin{figure}[t]
  \centering
  \includegraphics[trim=0cm 0cm 0cm 0cm, clip=true,width=0.5\textwidth]{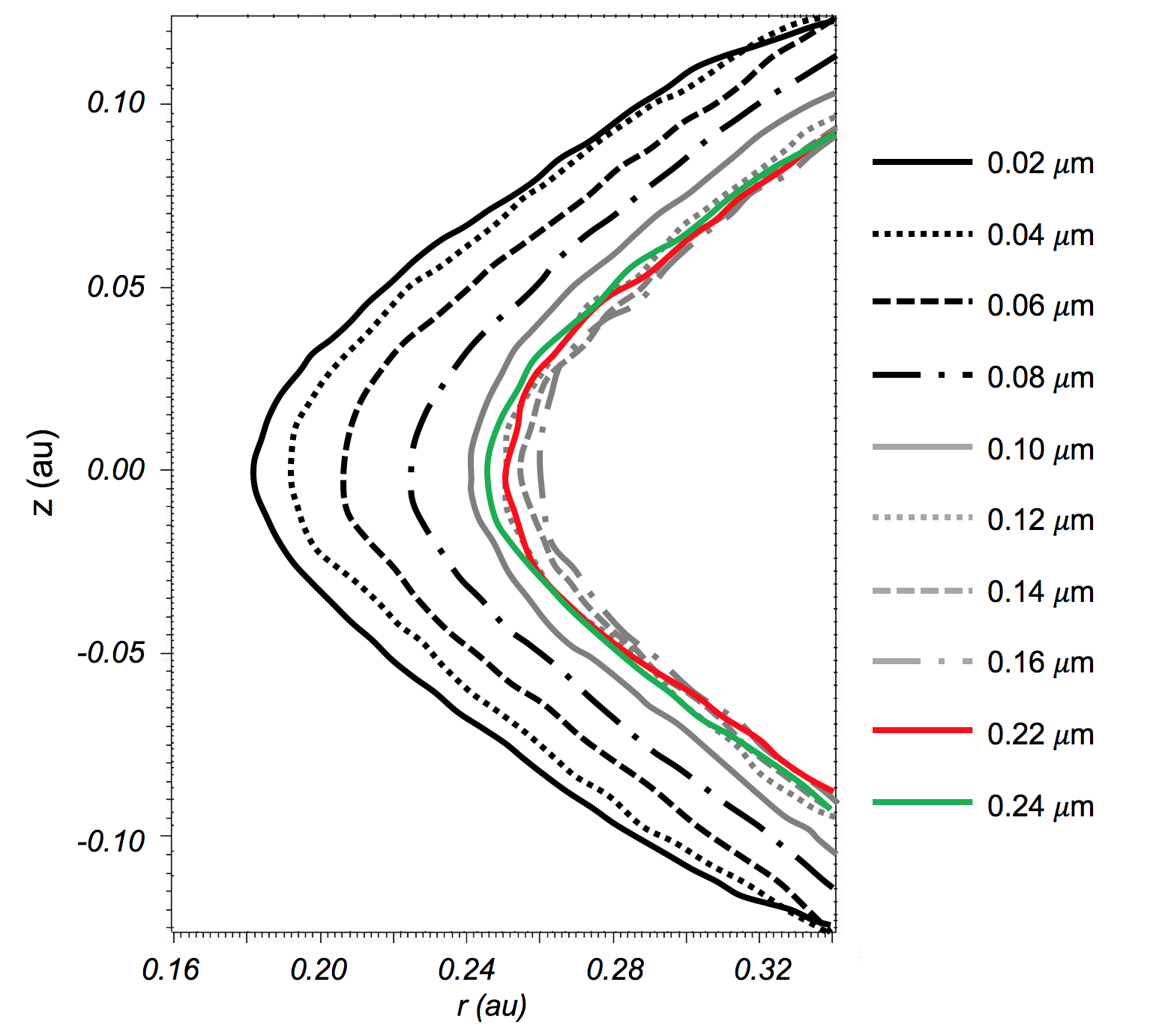}
  \caption{Inner rim shape and location for models computed with $a_{\rm{max}}$ between $0.02\,\mu$m and $0.16\,\mu$m (various black and grey lines - see the key on the right for details) compared to models with $a_{\rm{max}}=0.22\,\mu$m (red line) and $a_{\rm{max}}=0.24\,\mu$m (green line). All models were run with $h_{\rm{0}}=9\,$au and $\beta=1.02$, the same as in our best-fit model (see Section~\ref{sec:results}).}
  \label{fig:inner_rim}
\end{figure}

We uncovered surprising behavior of the dependence of the location of $R_{\rm{sub}}$ and the grain size when $a_{\rm{max}}<0.16\,\mu$m. In Fig.~\ref{fig:inner_rim}, we show the location and shape of the inner edge of the rim for $a_{\rm{max}}$ between $0.02\,\mu$m and $0.16\,\mu$m and compare these to models with $a_{\rm{max}}=0.22\,\mu$m and $a_{\rm{max}}=0.24\,\mu$m which behave as expected. Rim shapes for models with $a_{\rm{max}}=0.18\,\mu$m and $a_{\rm{max}}=0.20\,\mu$m were indistinguishable from the model with $a_{\rm{max}}=0.16\,\mu$m and are thus not shown in the plot. Between $0.02\,\mu$m and $\approx0.16\,\mu$m, the inner rim location moves further from the star with increasing $a_{\rm{max}}$, opposite to what is expected and which has been reported for similar studies of hotter stars \citep{Isella05, Davies18}. Models including grains larger than $\approx0.16\,\mu$m maintain the behavior which we expect to see: the inner rim location moves closer to the star with increasing $a_{\rm{max}}$. Further investigation into the reasons for this are outside the scope of this paper and are deferred to future study (Davies \& Harries 2020, in preparation).

\vspace{1cm}


\bibliography{rytau}

\begin{thebibliography}{}
\expandafter\ifx\csname natexlab\endcsname\relax\def\natexlab#1{#1}\fi
\providecommand{\url}[1]{\href{#1}{#1}}
\providecommand{\dodoi}[1]{doi:~\href{http://doi.org/#1}{\nolinkurl{#1}}}
\providecommand{\doeprint}[1]{\href{http://ascl.net/#1}{\nolinkurl{http://ascl.net/#1}}}
\providecommand{\doarXiv}[1]{\href{https://arxiv.org/abs/#1}{\nolinkurl{https://arxiv.org/abs/#1}}}

\bibitem[{{Abrahamyan} {et~al.}(2015){Abrahamyan}, {Mickaelian}, \&
  {Knyazyan}}]{Abrahamyan15}
{Abrahamyan}, H.~V., {Mickaelian}, A.~M., \& {Knyazyan}, A.~V. 2015, Astronomy
  and Computing, 10, 99, \dodoi{10.1016/j.ascom.2014.12.002}

\bibitem[{{Agra-Amboage} {et~al.}(2009){Agra-Amboage}, {Dougados}, {Cabrit},
  {Garcia}, \& {Ferruit}}]{Agra09}
{Agra-Amboage}, V., {Dougados}, C., {Cabrit}, S., {Garcia}, P.~J.~V., \&
  {Ferruit}, P. 2009, \aap, 493, 1029, \dodoi{10.1051/0004-6361:200810025}

\bibitem[{{Akeson} {et~al.}(2000){Akeson}, {Ciardi}, {van Belle},
  {Creech-Eakman}, \& {Lada}}]{Akeson00}
{Akeson}, R.~L., {Ciardi}, D.~R., {van Belle}, G.~T., {Creech-Eakman}, M.~J.,
  \& {Lada}, E.~A. 2000, \apj, 543, 313, \dodoi{10.1086/317111}

\bibitem[{{Akeson} {et~al.}(2005){Akeson}, {Walker}, {Wood}, {Eisner}, {Scire},
  {Penprase}, {Ciardi}, {van Belle}, {Whitney}, \& {Bjorkman}}]{Akeson05}
{Akeson}, R.~L., {Walker}, C.~H., {Wood}, K., {et~al.} 2005, \apj, 622, 440,
  \dodoi{10.1086/427770}

\bibitem[{{Altenhoff} {et~al.}(1994){Altenhoff}, {Thum}, \&
  {Wendker}}]{Altenhoff94}
{Altenhoff}, W.~J., {Thum}, C., \& {Wendker}, H.~J. 1994, \aap, 281, 161

\bibitem[{{Andrews} {et~al.}(2013){Andrews}, {Rosenfeld}, {Kraus}, \&
  {Wilner}}]{Andrews13}
{Andrews}, S.~M., {Rosenfeld}, K.~A., {Kraus}, A.~L., \& {Wilner}, D.~J. 2013,
  \apj, 771, 129, \dodoi{10.1088/0004-637X/771/2/129}

\bibitem[{{Andrews} \& {Williams}(2005)}]{Andrews05}
{Andrews}, S.~M., \& {Williams}, J.~P. 2005, \apj, 631, 1134,
  \dodoi{10.1086/432712}

\bibitem[{{Anugu} {et~al.}(2018){Anugu}, {Le Bouquin}, {Monnier}, {Kraus},
  {Ennis}, {Lanthermann}, {Setterholm}, {Davies}, {ten Brummelaar}, {Haidar},
  {Dubravec}, \& {Peters}}]{Anugu18}
{Anugu}, N., {Le Bouquin}, J.-B., {Monnier}, J.~D., {et~al.} 2018, in Society
  of Photo-Optical Instrumentation Engineers (SPIE) Conference Series, Vol.
  10701, \procspie, 1070124, \dodoi{10.1117/12.2313036}

\bibitem[{{Astropy Collaboration} {et~al.}(2013){Astropy Collaboration},
  {Robitaille}, {Tollerud}, {Greenfield}, {Droettboom}, {Bray}, {Aldcroft},
  {Davis}, {Ginsburg}, {Price-Whelan}, {Kerzendorf}, {Conley}, {Crighton},
  {Barbary}, {Muna}, {Ferguson}, {Grollier}, {Parikh}, {Nair}, {Unther},
  {Deil}, {Woillez}, {Conseil}, {Kramer}, {Turner}, {Singer}, {Fox}, {Weaver},
  {Zabalza}, {Edwards}, {Azalee Bostroem}, {Burke}, {Casey}, {Crawford},
  {Dencheva}, {Ely}, {Jenness}, {Labrie}, {Lim}, {Pierfederici}, {Pontzen},
  {Ptak}, {Refsdal}, {Servillat}, \& {Streicher}}]{astropy}
{Astropy Collaboration}, {Robitaille}, T.~P., {Tollerud}, E.~J., {et~al.} 2013,
  \aap, 558, A33, \dodoi{10.1051/0004-6361/201322068}

\bibitem[{{Babina} {et~al.}(2016){Babina}, {Artemenko}, \& {Petrov}}]{Babina16}
{Babina}, E.~V., {Artemenko}, S.~A., \& {Petrov}, P.~P. 2016, Astronomy
  Letters, 42, 193, \dodoi{10.1134/S1063773716030014}

\bibitem[{{Bailer-Jones} {et~al.}(2018){Bailer-Jones}, {Rybizki}, {Fouesneau},
  {Mantelet}, \& {Andrae}}]{Bailer18}
{Bailer-Jones}, C.~A.~L., {Rybizki}, J., {Fouesneau}, M., {Mantelet}, G., \&
  {Andrae}, R. 2018, \aj, 156, 58, \dodoi{10.3847/1538-3881/aacb21}

\bibitem[{{Beckwith} \& {Sargent}(1991)}]{Beckwith91}
{Beckwith}, S.~V.~W., \& {Sargent}, A.~I. 1991, \apj, 381, 250,
  \dodoi{10.1086/170646}

\bibitem[{{Bodman} {et~al.}(2017){Bodman}, {Quillen}, {Ansdell}, {Hippke},
  {Boyajian}, {Mamajek}, {Blackman}, {Rizzuto}, \& {Kastner}}]{Bodman17}
{Bodman}, E.~H.~L., {Quillen}, A.~C., {Ansdell}, M., {et~al.} 2017, \mnras,
  470, 202, \dodoi{10.1093/mnras/stx1034}

\bibitem[{{Bonneau} {et~al.}(2011){Bonneau}, {Delfosse}, {Mourard}, {Lafrasse},
  {Mella}, {Cetre}, {Clausse}, \& {Zins}}]{Bonneau11}
{Bonneau}, D., {Delfosse}, X., {Mourard}, D., {et~al.} 2011, \aap, 535, A53,
  \dodoi{10.1051/0004-6361/201015124}

\bibitem[{{Bonneau} {et~al.}(2006){Bonneau}, {Clausse}, {Delfosse}, {Mourard},
  {Cetre}, {Chelli}, {Cruzal{\`e}bes}, {Duvert}, \& {Zins}}]{Bonneau06}
{Bonneau}, D., {Clausse}, J.-M., {Delfosse}, X., {et~al.} 2006, \aap, 456, 789,
  \dodoi{10.1051/0004-6361:20054469}

\bibitem[{{Calvet} {et~al.}(2004){Calvet}, {Muzerolle}, {Brice{\~n}o},
  {Hern{\'a}ndez}, {Hartmann}, {Saucedo}, \& {Gordon}}]{Calvet04}
{Calvet}, N., {Muzerolle}, J., {Brice{\~n}o}, C., {et~al.} 2004, \aj, 128,
  1294, \dodoi{10.1086/422733}

\bibitem[{{Cieza} {et~al.}(2009){Cieza}, {Padgett}, {Allen}, {McCabe},
  {Brooke}, {Carey}, {Chapman}, {Fukagawa}, {Huard}, {Noriga-Crespo},
  {Peterson}, \& {Rebull}}]{Cieza09}
{Cieza}, L.~A., {Padgett}, D.~L., {Allen}, L.~E., {et~al.} 2009, \apjl, 696,
  L84, \dodoi{10.1088/0004-637X/696/1/L84}

\bibitem[{{Colavita} {et~al.}(1999){Colavita}, {Wallace}, {Hines}, {Gursel},
  {Malbet}, {Palmer}, {Pan}, {Shao}, {Yu}, {Boden}, {Dumont}, {Gubler},
  {Koresko}, {Kulkarni}, {Lane}, {Mobley}, \& {van Belle}}]{Colavita99}
{Colavita}, M.~M., {Wallace}, J.~K., {Hines}, B.~E., {et~al.} 1999, \apj, 510,
  505, \dodoi{10.1086/306579}

\bibitem[{{Colavita} {et~al.}(2013){Colavita}, {Wizinowich}, {Akeson},
  {Ragland}, {Woillez}, {Millan-Gabet}, {Serabyn}, {Abajian}, {Acton},
  {Appleby}, {Beletic}, {Beichman}, {Bell}, {Berkey}, {Berlin}, {Boden},
  {Booth}, {Boutell}, {Chaffee}, {Chan}, {Chin}, {Chock}, {Cohen}, {Cooper},
  {Crawford}, {Creech-Eakman}, {Dahl}, {Eychaner}, {Fanson}, {Felizardo},
  {Garcia-Gathright}, {Gathright}, {Hardy}, {Henderson}, {Herstein}, {Hess},
  {Hovland}, {Hrynevych}, {Johansson}, {Johnson}, {Kelley}, {Kendrick},
  {Koresko}, {Kurpis}, {Le Mignant}, {Lewis}, {Ligon}, {Lupton}, {McBride},
  {Medeiros}, {Mennesson}, {Moore}, {Morrison}, {Nance}, {Neyman}, {Niessner},
  {Paine}, {Palmer}, {Panteleeva}, {Papin}, {Parvin}, {Reder}, {Rudeen},
  {Saloga}, {Sargent}, {Shao}, {Smith}, {Smythe}, {Stomski}, {Summers},
  {Swain}, {Swanson}, {Thompson}, {Tsubota}, {Tumminello}, {Tyau}, {van Belle},
  {Vasisht}, {Vause}, {Vescelus}, {Walker}, {Wallace}, {Wehmeier}, \&
  {Wetherell}}]{Colavita13}
{Colavita}, M.~M., {Wizinowich}, P.~L., {Akeson}, R.~L., {et~al.} 2013, \pasp,
  125, 1226, \dodoi{10.1086/673475}

\bibitem[{{Davies} {et~al.}(2018){Davies}, {Kraus}, {Harries}, {Kreplin},
  {Monnier}, {Labdon}, {Kloppenborg}, {Acreman}, {Baron}, {Millan-Gabet},
  {Sturmann}, {Sturmann}, \& {Ten Brummelaar}}]{Davies18}
{Davies}, C.~L., {Kraus}, S., {Harries}, T.~J., {et~al.} 2018, \apj, 866, 23,
  \dodoi{10.3847/1538-4357/aade51}

\bibitem[{{Dionatos} {et~al.}(2019){Dionatos}, {Woitke}, {G{\"u}del},
  {Degroote}, {Liebhart}, {Anthonioz}, {Antonellini}, {Baldovin-Saavedra},
  {Carmona}, {Dominik}, {Greaves}, {Ilee}, {Kamp}, {M{\'e}nard}, {Min},
  {Pinte}, {Rab}, {Rigon}, {Thi}, \& {Waters}}]{Dionatos19}
{Dionatos}, O., {Woitke}, P., {G{\"u}del}, M., {et~al.} 2019, \aap, 625, A66,
  \dodoi{10.1051/0004-6361/201832860}

\bibitem[{{Draine}(2003)}]{Draine03}
{Draine}, B.~T. 2003, \apj, 598, 1026, \dodoi{10.1086/379123}

\bibitem[{{Duvert} {et~al.}(2017){Duvert}, {Young}, \& {Hummel}}]{Duvert17}
{Duvert}, G., {Young}, J., \& {Hummel}, C.~A. 2017, \aap, 597, A8,
  \dodoi{10.1051/0004-6361/201526405}

\bibitem[{{Elias}(1978)}]{Elias78}
{Elias}, J.~H. 1978, \apj, 224, 857, \dodoi{10.1086/156436}

\bibitem[{{Espaillat} {et~al.}(2011){Espaillat}, {Furlan}, {D'Alessio},
  {Sargent}, {Nagel}, {Calvet}, {Watson}, \& {Muzerolle}}]{Espaillat11}
{Espaillat}, C., {Furlan}, E., {D'Alessio}, P., {et~al.} 2011, \apj, 728, 49,
  \dodoi{10.1088/0004-637X/728/1/49}

\bibitem[{{Furlan} {et~al.}(2009){Furlan}, {Watson}, {McClure}, {Manoj},
  {Espaillat}, {D'Alessio}, {Calvet}, {Kim}, {Sargent}, {Forrest}, \&
  {Hartmann}}]{Furlan09}
{Furlan}, E., {Watson}, D.~M., {McClure}, M.~K., {et~al.} 2009, \apj, 703,
  1964, \dodoi{10.1088/0004-637X/703/2/1964}

\bibitem[{{Gaia Collaboration} {et~al.}(2016){Gaia Collaboration}, {Prusti},
  {de Bruijne}, {Brown}, {Vallenari}, {Babusiaux}, {Bailer-Jones}, {Bastian},
  {Biermann}, {Evans}, \& et~al.}]{Gaia16}
{Gaia Collaboration}, {Prusti}, T., {de Bruijne}, J.~H.~J., {et~al.} 2016,
  \aap, 595, A1, \dodoi{10.1051/0004-6361/201629272}

\bibitem[{{Gaia Collaboration} {et~al.}(2018){Gaia Collaboration}, {Brown},
  {Vallenari}, {Prusti}, {de Bruijne}, {Babusiaux}, {Bailer-Jones}, {Biermann},
  {Evans}, {Eyer}, {Jansen}, {Jordi}, {Klioner}, {Lammers}, {Lindegren},
  {Luri}, {Mignard}, {Panem}, {Pourbaix}, {Randich}, {Sartoretti}, {Siddiqui},
  {Soubiran}, {van Leeuwen}, {Walton}, {Arenou}, {Bastian}, {Cropper},
  {Drimmel}, {Katz}, {Lattanzi}, {Bakker}, {Cacciari}, {Casta{\~n}eda},
  {Chaoul}, {Cheek}, {De Angeli}, {Fabricius}, {Guerra}, {Holl}, {Masana},
  {Messineo}, {Mowlavi}, {Nienartowicz}, {Panuzzo}, {Portell}, {Riello},
  {Seabroke}, {Tanga}, {Th{\'e}venin}, {Gracia-Abril}, {Comoretto},
  {Garcia-Reinaldos}, {Teyssier}, {Altmann}, {Andrae}, {Audard},
  {Bellas-Velidis}, {Benson}, {Berthier}, {Blomme}, {Burgess}, {Busso},
  {Carry}, {Cellino}, {Clementini}, {Clotet}, {Creevey}, {Davidson}, {De
  Ridder}, {Delchambre}, {Dell'Oro}, {Ducourant},
  {Fern{\'a}ndez-Hern{\'a}ndez}, {Fouesneau}, {Fr{\'e}mat}, {Galluccio},
  {Garc{\'\i}a-Torres}, {Gonz{\'a}lez-N{\'u}{\~n}ez}, {Gonz{\'a}lez-Vidal},
  {Gosset}, {Guy}, {Halbwachs}, {Hambly}, {Harrison}, {Hern{\'a}ndez},
  {Hestroffer}, {Hodgkin}, {Hutton}, {Jasniewicz}, {Jean-Antoine-Piccolo},
  {Jordan}, {Korn}, {Krone-Martins}, {Lanzafame}, {Lebzelter}, {L{\"o}ffler},
  {Manteiga}, {Marrese}, {Mart{\'\i}n-Fleitas}, {Moitinho}, {Mora}, {Muinonen},
  {Osinde}, {Pancino}, {Pauwels}, {Petit}, {Recio-Blanco}, {Richards},
  {Rimoldini}, {Robin}, {Sarro}, {Siopis}, {Smith}, {Sozzetti}, {S{\"u}veges},
  {Torra}, {van Reeven}, {Abbas}, {Abreu Aramburu}, {Accart}, {Aerts},
  {Altavilla}, {{\'A}lvarez}, {Alvarez}, {Alves}, {Anderson}, {Andrei},
  {Anglada Varela}, {Antiche}, {Antoja}, {Arcay}, {Astraatmadja}, {Bach},
  {Baker}, {Balaguer-N{\'u}{\~n}ez}, {Balm}, {Barache}, {Barata}, {Barbato},
  {Barblan}, {Barklem}, {Barrado}, {Barros}, {Barstow}, {Bartholom{\'e}
  Mu{\~n}oz}, {Bassilana}, {Becciani}, {Bellazzini}, {Berihuete}, {Bertone},
  {Bianchi}, {Bienaym{\'e}}, {Blanco-Cuaresma}, {Boch}, {Boeche}, {Bombrun},
  {Borrachero}, {Bossini}, {Bouquillon}, {Bourda}, {Bragaglia}, {Bramante},
  {Breddels}, {Bressan}, {Brouillet}, {Br{\"u}semeister}, {Brugaletta},
  {Bucciarelli}, {Burlacu}, {Busonero}, {Butkevich}, {Buzzi}, {Caffau},
  {Cancelliere}, {Cannizzaro}, {Cantat-Gaudin}, {Carballo}, {Carlucci},
  {Carrasco}, {Casamiquela}, {Castellani}, {Castro-Ginard}, {Charlot},
  {Chemin}, {Chiavassa}, {Cocozza}, {Costigan}, {Cowell}, {Crifo}, {Crosta},
  {Crowley}, {Cuypers}, {Dafonte}, {Damerdji}, {Dapergolas}, {David}, {David},
  {de Laverny}, {De Luise}, {De March}, {de Martino}, {de Souza}, {de Torres},
  {Debosscher}, {del Pozo}, {Delbo}, {Delgado}, {Delgado}, {Di Matteo},
  {Diakite}, {Diener}, {Distefano}, {Dolding}, {Drazinos}, {Dur{\'a}n},
  {Edvardsson}, {Enke}, {Eriksson}, {Esquej}, {Eynard Bontemps}, {Fabre},
  {Fabrizio}, {Faigler}, {Falc{\~a}o}, {Farr{\`a}s Casas}, {Federici},
  {Fedorets}, {Fernique}, {Figueras}, {Filippi}, {Findeisen}, {Fonti},
  {Fraile}, {Fraser}, {Fr{\'e}zouls}, {Gai}, {Galleti}, {Garabato},
  {Garc{\'\i}a-Sedano}, {Garofalo}, {Garralda}, {Gavel}, {Gavras}, {Gerssen},
  {Geyer}, {Giacobbe}, {Gilmore}, {Girona}, {Giuffrida}, {Glass}, {Gomes},
  {Granvik}, {Gueguen}, {Guerrier}, {Guiraud}, {Guti{\'e}rrez-S{\'a}nchez},
  {Haigron}, {Hatzidimitriou}, {Hauser}, {Haywood}, {Heiter}, {Helmi}, {Heu},
  {Hilger}, {Hobbs}, {Hofmann}, {Holland}, {Huckle}, {Hypki}, {Icardi},
  {Jan{\ss}en}, {Jevardat de Fombelle}, {Jonker}, {Juh{\'a}sz}, {Julbe},
  {Karampelas}, {Kewley}, {Klar}, {Kochoska}, {Kohley}, {Kolenberg},
  {Kontizas}, {Kontizas}, {Koposov}, {Kordopatis}, {Kostrzewa-Rutkowska},
  {Koubsky}, {Lambert}, {Lanza}, {Lasne}, {Lavigne}, {Le Fustec}, {Le
  Poncin-Lafitte}, {Lebreton}, {Leccia}, {Leclerc}, {Lecoeur-Taibi},
  {Lenhardt}, {Leroux}, {Liao}, {Licata}, {Lindstr{\o}m}, {Lister}, {Livanou},
  {Lobel}, {L{\'o}pez}, {Managau}, {Mann}, {Mantelet}, {Marchal}, {Marchant},
  {Marconi}, {Marinoni}, {Marschalk{\'o}}, {Marshall}, {Martino}, {Marton},
  {Mary}, {Massari}, {Matijevi{\v{c}}}, {Mazeh}, {McMillan}, {Messina},
  {Michalik}, {Millar}, {Molina}, {Molinaro}, {Moln{\'a}r}, {Montegriffo},
  {Mor}, {Morbidelli}, {Morel}, {Morris}, {Mulone}, {Muraveva}, {Musella},
  {Nelemans}, {Nicastro}, {Noval}, {O'Mullane}, {Ord{\'e}novic},
  {Ord{\'o}{\~n}ez-Blanco}, {Osborne}, {Pagani}, {Pagano}, {Pailler},
  {Palacin}, {Palaversa}, {Panahi}, {Pawlak}, {Piersimoni}, {Pineau}, {Plachy},
  {Plum}, {Poggio}, {Poujoulet}, {Pr{\v{s}}a}, {Pulone}, {Racero}, {Ragaini},
  {Rambaux}, {Ramos-Lerate}, {Regibo}, {Reyl{\'e}}, {Riclet}, {Ripepi}, {Riva},
  {Rivard}, {Rixon}, {Roegiers}, {Roelens}, {Romero-G{\'o}mez}, {Rowell},
  {Royer}, {Ruiz-Dern}, {Sadowski}, {Sagrist{\`a} Sell{\'e}s}, {Sahlmann},
  {Salgado}, {Salguero}, {Sanna}, {Santana-Ros}, {Sarasso}, {Savietto},
  {Schultheis}, {Sciacca}, {Segol}, {Segovia}, {S{\'e}gransan}, {Shih},
  {Siltala}, {Silva}, {Smart}, {Smith}, {Solano}, {Solitro}, {Sordo}, {Soria
  Nieto}, {Souchay}, {Spagna}, {Spoto}, {Stampa}, {Steele},
  {Steidelm{\"u}ller}, {Stephenson}, {Stoev}, {Suess}, {Surdej}, {Szabados},
  {Szegedi-Elek}, {Tapiador}, {Taris}, {Tauran}, {Taylor}, {Teixeira},
  {Terrett}, {Teyssand ier}, {Thuillot}, {Titarenko}, {Torra Clotet}, {Turon},
  {Ulla}, {Utrilla}, {Uzzi}, {Vaillant}, {Valentini}, {Valette}, {van Elteren},
  {Van Hemelryck}, {van Leeuwen}, {Vaschetto}, {Vecchiato}, {Veljanoski},
  {Viala}, {Vicente}, {Vogt}, {von Essen}, {Voss}, {Votruba}, {Voutsinas},
  {Walmsley}, {Weiler}, {Wertz}, {Wevers}, {Wyrzykowski}, {Yoldas},
  {{\v{Z}}erjal}, {Ziaeepour}, {Zorec}, {Zschocke}, {Zucker}, {Zurbach}, \&
  {Zwitter}}]{Gaia18}
{Gaia Collaboration}, {Brown}, A.~G.~A., {Vallenari}, A., {et~al.} 2018, \aap,
  616, A1, \dodoi{10.1051/0004-6361/201833051}

\bibitem[{{Galli} {et~al.}(2018){Galli}, {Loinard}, {Ortiz-L{\'e}on},
  {Kounkel}, {Dzib}, {Mioduszewski}, {Rodr{\'{\i}}guez}, {Hartmann},
  {Teixeira}, {Torres}, {Rivera}, {Boden}, {Evans}, {Brice{\~n}o}, {Tobin}, \&
  {Heyer}}]{Galli18}
{Galli}, P.~A.~B., {Loinard}, L., {Ortiz-L{\'e}on}, G.~N., {et~al.} 2018, \apj,
  859, 33, \dodoi{10.3847/1538-4357/aabf91}

\bibitem[{{Garufi} {et~al.}(2019){Garufi}, {Podio}, {Bacciotti}, {Antoniucci},
  {Boccaletti}, {Codella}, {Dougados}, {M{\'e}nard}, {Mesa}, {Meyer}, {Nisini},
  {Schmid}, {Stolker}, {Baudino}, {Biller}, {Bonavita}, {Bonnefoy},
  {Cantalloube}, {Chauvin}, {Cheetham}, {Desidera}, {D'Orazi}, {Feldt},
  {Galicher}, {Grandjean}, {Gratton}, {Hagelberg}, {Lagrange}, {Langlois},
  {Lannier}, {Lazzoni}, {Maire}, {Perrot}, {Rickman}, {Schmidt}, {Vigan},
  {Zurlo}, {Delboulb{\'e}}, {Le Mignant}, {Fantinel}, {M{\"o}ller-Nilsson},
  {Weber}, \& {Sauvage}}]{Garufi19}
{Garufi}, A., {Podio}, L., {Bacciotti}, F., {et~al.} 2019, \aap, 628, A68,
  \dodoi{10.1051/0004-6361/201935546}

\bibitem[{{Grankin} {et~al.}(2007){Grankin}, {Melnikov}, {Bouvier}, {Herbst},
  \& {Shevchenko}}]{Grankin07}
{Grankin}, K.~N., {Melnikov}, S.~Y., {Bouvier}, J., {Herbst}, W., \&
  {Shevchenko}, V.~S. 2007, \aap, 461, 183, \dodoi{10.1051/0004-6361:20065489}

\bibitem[{{Gravity Collaboration} {et~al.}(2017){Gravity Collaboration},
  {Abuter}, {Accardo}, {Amorim}, {Anugu}, {{\'A}vila}, {Azouaoui}, {Benisty},
  {Berger}, {Blind}, {Bonnet}, {Bourget}, {Brandner}, {Brast}, {Buron},
  {Burtscher}, {Cassaing}, {Chapron}, {Choquet}, {Cl{\'e}net}, {Collin},
  {Coud{\'e} Du Foresto}, {de Wit}, {de Zeeuw}, {Deen},
  {Delplancke-Str{\"o}bele}, {Dembet}, {Derie}, {Dexter}, {Duvert}, {Ebert},
  {Eckart}, {Eisenhauer}, {Esselborn}, {F{\'e}dou}, {Finger}, {Garcia}, {Garcia
  Dabo}, {Garcia Lopez}, {Gendron}, {Genzel}, {Gillessen}, {Gonte}, {Gordo},
  {Grould}, {Gr{\"o}zinger}, {Guieu}, {Haguenauer}, {Hans}, {Haubois}, {Haug},
  {Haussmann}, {Henning}, {Hippler}, {Horrobin}, {Huber}, {Hubert}, {Hubin},
  {Hummel}, {Jakob}, {Janssen}, {Jochum}, {Jocou}, {Kaufer}, {Kellner},
  {Kendrew}, {Kern}, {Kervella}, {Kiekebusch}, {Klein}, {Kok}, {Kolb}, {Kulas},
  {Lacour}, {Lapeyr{\`e}re}, {Lazareff}, {Le Bouquin}, {L{\`e}na}, {Lenzen},
  {L{\'e}v{\^e}que}, {Lippa}, {Magnard}, {Mehrgan}, {Mellein}, {M{\'e}rand},
  {Moreno-Ventas}, {Moulin}, {M{\"u}ller}, {M{\"u}ller}, {Neumann}, {Oberti},
  {Ott}, {Pallanca}, {Panduro}, {Pasquini}, {Paumard}, {Percheron}, {Perraut},
  {Perrin}, {Pfl{\"u}ger}, {Pfuhl}, {Phan Duc}, {Plewa}, {Popovic}, {Rabien},
  {Ram{\'{\i}}rez}, {Ramos}, {Rau}, {Riquelme}, {Rohloff}, {Rousset},
  {Sanchez-Bermudez}, {Scheithauer}, {Sch{\"o}ller}, {Schuhler}, {Spyromilio},
  {Straubmeier}, {Sturm}, {Suarez}, {Tristram}, {Ventura}, {Vincent},
  {Waisberg}, {Wank}, {Weber}, {Wieprecht}, {Wiest}, {Wiezorrek}, {Wittkowski},
  {Woillez}, {Wolff}, {Yazici}, {Ziegler}, \& {Zins}}]{gravity}
{Gravity Collaboration}, {Abuter}, R., {Accardo}, M., {et~al.} 2017, \aap, 602,
  A94, \dodoi{10.1051/0004-6361/201730838}

\bibitem[{{GRAVITY Collaboration} {et~al.}(2019){GRAVITY Collaboration},
  {Perraut}, {Labadie}, {Lazareff}, {Klarmann}, {Segura-Cox}, {Benisty},
  {Bouvier}, {Brandner}, {Caratti O Garatti}, {Caselli}, {Dougados}, {Garcia},
  {Garcia-Lopez}, {Kendrew}, {Koutoulaki}, {Kervella}, {Lin}, {Pineda},
  {Sanchez-Bermudez}, {van Dishoeck}, {Abuter}, {Amorim}, {Berger}, {Bonnet},
  {Buron}, {Cantalloube}, {Cl{\'e}net}, {Coud{\'e} Du Foresto}, {Dexter}, {de
  Zeeuw}, {Duvert}, {Eckart}, {Eisenhauer}, {Eupen}, {Gao}, {Gendron},
  {Genzel}, {Gillessen}, {Gordo}, {Grellmann}, {Haubois}, {Haussmann},
  {Henning}, {Hippler}, {Horrobin}, {Hubert}, {Jocou}, {Lacour}, {Le Bouquin},
  {L{\'e}na}, {M{\'e}rand}, {Ott}, {Paumard}, {Perrin}, {Pfuhl}, {Rabien},
  {Ray}, {Rau}, {Rousset}, {Scheithauer}, {Straub}, {Straubmeier}, {Sturm},
  {Vincent}, {Waisberg}, {Wank}, {Widmann}, {Wieprecht}, {Wiest}, {Wiezorrek},
  {Woillez}, \& {Yazici}}]{Gravity19}
{GRAVITY Collaboration}, {Perraut}, K., {Labadie}, L., {et~al.} 2019, \aap,
  632, A53, \dodoi{10.1051/0004-6361/201936403}

\bibitem[{{Gregory} {et~al.}(2016){Gregory}, {Donati}, \&
  {Hussain}}]{Gregory16}
{Gregory}, S.~G., {Donati}, J.-F., \& {Hussain}, G.~A.~J. 2016, ArXiv e-prints.
\newblock \doarXiv{1609.00273}

\bibitem[{{Harries}(2000)}]{Harries00}
{Harries}, T.~J. 2000, \mnras, 315, 722,
  \dodoi{10.1046/j.1365-8711.2000.03505.x}

\bibitem[{{Harries} {et~al.}(2019){Harries}, {Haworth}, {Acreman}, {Ali}, \&
  {Douglas}}]{Harries19}
{Harries}, T.~J., {Haworth}, T.~J., {Acreman}, D., {Ali}, A., \& {Douglas}, T.
  2019, Astronomy and Computing, 27, 63, \dodoi{10.1016/j.ascom.2019.03.002}

\bibitem[{{Haubois} {et~al.}(2014){Haubois}, {Bernaud}, {Mella}, {Duvert},
  {Benisty}, {B{\'e}rio}, {Bourges}, {Chelli}, {Chesneau}, {Lacour},
  {Lafrasse}, {Le Bouquin}, {Mourard}, {Nardetto}, \& {Olofsson}}]{Haubois14}
{Haubois}, X., {Bernaud}, P., {Mella}, G., {et~al.} 2014, in \procspie, Vol.
  9146, Optical and Infrared Interferometry IV, 91460O,
  \dodoi{10.1117/12.2056977}

\bibitem[{{Herczeg} \& {Hillenbrand}(2014)}]{Herczeg14}
{Herczeg}, G.~J., \& {Hillenbrand}, L.~A. 2014, \apj, 786, 97,
  \dodoi{10.1088/0004-637X/786/2/97}

\bibitem[{{Houck} {et~al.}(2004){Houck}, {Roellig}, {van Cleve}, {Forrest},
  {Herter}, {Lawrence}, {Matthews}, {Reitsema}, {Soifer}, {Watson}, {Weedman},
  {Huisjen}, {Troeltzsch}, {Barry}, {Bernard-Salas}, {Blacken}, {Brandl},
  {Charmandaris}, {Devost}, {Gull}, {Hall}, {Henderson}, {Higdon}, {Pirger},
  {Schoenwald}, {Sloan}, {Uchida}, {Appleton}, {Armus}, {Burgdorf},
  {Fajardo-Acosta}, {Grillmair}, {Ingalls}, {Morris}, \& {Teplitz}}]{Houck04}
{Houck}, J.~R., {Roellig}, T.~L., {van Cleve}, J., {et~al.} 2004, \apjs, 154,
  18, \dodoi{10.1086/423134}

\bibitem[{{Howard} {et~al.}(2013){Howard}, {Sandell}, {Vacca}, {Duch{\^e}ne},
  {Mathews}, {Augereau}, {Barrado}, {Dent}, {Eiroa}, {Grady}, {Kamp}, {Meeus},
  {M{\'e}nard}, {Pinte}, {Podio}, {Riviere-Marichalar}, {Roberge}, {Thi},
  {Vicente}, \& {Williams}}]{Howard13}
{Howard}, C.~D., {Sandell}, G., {Vacca}, W.~D., {et~al.} 2013, \apj, 776, 21,
  \dodoi{10.1088/0004-637X/776/1/21}

\bibitem[{Hunter(2007)}]{Hunter07}
Hunter, J.~D. 2007, Computing In Science \& Engineering, 9, 90

\bibitem[{{Isella} {et~al.}(2010){Isella}, {Carpenter}, \&
  {Sargent}}]{Isella10}
{Isella}, A., {Carpenter}, J.~M., \& {Sargent}, A.~I. 2010, \apj, 714, 1746,
  \dodoi{10.1088/0004-637X/714/2/1746}

\bibitem[{{Isella} \& {Natta}(2005)}]{Isella05}
{Isella}, A., \& {Natta}, A. 2005, \aap, 438, 899,
  \dodoi{10.1051/0004-6361:20052773}

\bibitem[{{Johnstone} {et~al.}(2014){Johnstone}, {Jardine}, {Gregory},
  {Donati}, \& {Hussain}}]{Johnstone14}
{Johnstone}, C.~P., {Jardine}, M., {Gregory}, S.~G., {Donati}, J.-F., \&
  {Hussain}, G. 2014, \mnras, 437, 3202, \dodoi{10.1093/mnras/stt2107}

\bibitem[{{Kama} {et~al.}(2009){Kama}, {Min}, \& {Dominik}}]{Kama09}
{Kama}, M., {Min}, M., \& {Dominik}, C. 2009, \aap, 506, 1199,
  \dodoi{10.1051/0004-6361/200912068}

\bibitem[{{Keane} {et~al.}(2014){Keane}, {Pascucci}, {Espaillat}, {Woitke},
  {Andrews}, {Kamp}, {Thi}, {Meeus}, \& {Dent}}]{Keane14}
{Keane}, J.~T., {Pascucci}, I., {Espaillat}, C., {et~al.} 2014, \apj, 787, 153,
  \dodoi{10.1088/0004-637X/787/2/153}

\bibitem[{{Kenyon} {et~al.}(1994){Kenyon}, {Dobrzycka}, \&
  {Hartmann}}]{Kenyon94}
{Kenyon}, S.~J., {Dobrzycka}, D., \& {Hartmann}, L. 1994, \aj, 108, 1872,
  \dodoi{10.1086/117200}

\bibitem[{{Kenyon} \& {Hartmann}(1995)}]{Kenyon95}
{Kenyon}, S.~J., \& {Hartmann}, L. 1995, \apjs, 101, 117,
  \dodoi{10.1086/192235}

\bibitem[{{Kesseli} {et~al.}(2016){Kesseli}, {Petkova}, {Wood}, {Whitney},
  {Hillenbrand}, {Gregory}, {Stauffer}, {Morales-Calderon}, {Rebull}, \&
  {Alencar}}]{Kesseli16}
{Kesseli}, A.~Y., {Petkova}, M.~A., {Wood}, K., {et~al.} 2016, \apj, 828, 42,
  \dodoi{10.3847/0004-637X/828/1/42}

\bibitem[{{Kitamura} {et~al.}(2002){Kitamura}, {Momose}, {Yokogawa}, {Kawabe},
  {Tamura}, \& {Ida}}]{Kitamura02}
{Kitamura}, Y., {Momose}, M., {Yokogawa}, S., {et~al.} 2002, \apj, 581, 357,
  \dodoi{10.1086/344223}

\bibitem[{{Kraus} {et~al.}(2018){Kraus}, {Monnier}, {Anugu}, {Le Bouquin},
  {Davies}, {Ennis}, {Labdon}, {Lanthermann}, {Setterholm}, \& {ten
  Brummelaar}}]{Kraus18}
{Kraus}, S., {Monnier}, J.~D., {Anugu}, N., {et~al.} 2018, in Society of
  Photo-Optical Instrumentation Engineers (SPIE) Conference Series, Vol. 10701,
  \procspie, 1070123, \dodoi{10.1117/12.2311706}

\bibitem[{{Kurucz}(1979)}]{Kurucz79}
{Kurucz}, R.~L. 1979, \apjs, 40, 1, \dodoi{10.1086/190589}

\bibitem[{{Labdon} {et~al.}(2019){Labdon}, {Kraus}, {Davies}, {Kreplin},
  {Kluska}, {Harries}, {Monnier}, {ten Brummelaar}, {Baron}, {Millan-Gabet},
  {Kloppenborg}, {Eisner}, {Sturmann}, \& {Sturmann}}]{Labdon19}
{Labdon}, A., {Kraus}, S., {Davies}, C.~L., {et~al.} 2019, \aap, 627, A36,
  \dodoi{10.1051/0004-6361/201935331}

\bibitem[{{Lazareff} {et~al.}(2017){Lazareff}, {Berger}, {Kluska}, {Le
  Bouquin}, {Benisty}, {Malbet}, {Koen}, {Pinte}, {Thi}, {Absil}, {Baron},
  {Delboulb{\'e}}, {Duvert}, {Isella}, {Jocou}, {Juhasz}, {Kraus}, {Lachaume},
  {M{\'e}nard}, {Millan-Gabet}, {Monnier}, {Moulin}, {Perraut}, {Rochat},
  {Soulez}, {Tallon}, {Thi{\'e}baut}, {Traub}, \& {Zins}}]{Lazareff17}
{Lazareff}, B., {Berger}, J.-P., {Kluska}, J., {et~al.} 2017, \aap, 599, A85,
  \dodoi{10.1051/0004-6361/201629305}

\bibitem[{{Lebouteiller} {et~al.}(2011){Lebouteiller}, {Barry}, {Spoon},
  {Bernard-Salas}, {Sloan}, {Houck}, \& {Weedman}}]{Lebouteiller11}
{Lebouteiller}, V., {Barry}, D.~J., {Spoon}, H.~W.~W., {et~al.} 2011, \apjs,
  196, 8, \dodoi{10.1088/0067-0049/196/1/8}

\bibitem[{{Long} {et~al.}(2018){Long}, {Pinilla}, {Herczeg}, {Harsono},
  {Dipierro}, {Pascucci}, {Hendler}, {Tazzari}, {Ragusa}, {Salyk}, {Edwards},
  {Lodato}, {van de Plas}, {Johnstone}, {Liu}, {Boehler}, {Cabrit}, {Manara},
  {Menard}, {Mulders}, {Nisini}, {Fischer}, {Rigliaco}, {Banzatti}, {Avenhaus},
  \& {Gully-Santiago}}]{Long18}
{Long}, F., {Pinilla}, P., {Herczeg}, G.~J., {et~al.} 2018, \apj, 869, 17,
  \dodoi{10.3847/1538-4357/aae8e1}

\bibitem[{{Long} {et~al.}(2019){Long}, {Herczeg}, {Harsono}, {Pinilla},
  {Tazzari}, {Manara}, {Pascucci}, {Cabrit}, {Nisini}, {Johnstone}, {Edwards},
  {Salyk}, {Menard}, {Lodato}, {Boehler}, {Mace}, {Liu}, {Mulders}, {Hendler},
  {Ragusa}, {Fischer}, {Banzatti}, {Rigliaco}, {van de Plas}, {Dipierro},
  {Gully-Santiago}, \& {Lopez-Valdivia}}]{Long19}
{Long}, F., {Herczeg}, G.~J., {Harsono}, D., {et~al.} 2019, \apj, 882, 49,
  \dodoi{10.3847/1538-4357/ab2d2d}

\bibitem[{{Long} {et~al.}(2005){Long}, {Romanova}, \& {Lovelace}}]{Long05}
{Long}, M., {Romanova}, M.~M., \& {Lovelace}, R.~V.~E. 2005, \apj, 634, 1214,
  \dodoi{10.1086/497000}

\bibitem[{{Lopez} {et~al.}(2014){Lopez}, {Lagarde}, {Jaffe}, {Petrov},
  {Sch{\"o}ller}, {Antonelli}, {Beckmann}, {Berio}, {Bettonvil}, {Glindemann},
  {Gonzalez}, {Graser}, {Hofmann}, {Millour}, {Robbe-Dubois}, {Venema}, {Wolf},
  {Henning}, {Lanz}, {Weigelt}, {Agocs}, {Bailet}, {Bresson}, {Bristow},
  {Dugu{\'e}}, {Heininger}, {Kroes}, {Laun}, {Lehmitz}, {Neumann}, {Augereau},
  {Avila}, {Behrend}, {van Belle}, {Berger}, {van Boekel}, {Bonhomme},
  {Bourget}, {Brast}, {Clausse}, {Connot}, {Conzelmann}, {Cruzal{\`e}bes},
  {Csepany}, {Danchi}, {Delbo}, {Delplancke}, {Dominik}, {van Duin}, {Elswijk},
  {Fantei}, {Finger}, {Gabasch}, {Gay}, {Girard}, {Girault}, {Gitton},
  {Glazenborg}, {Gont{\'e}}, {Guitton}, {Guniat}, {De Haan}, {Haguenauer},
  {Hanenburg}, {Hogerheijde}, {ter Horst}, {Hron}, {Hugues}, {Hummel},
  {Idserda}, {Ives}, {Jakob}, {Jasko}, {Jolley}, {Kiraly}, {K{\"o}hler},
  {Kragt}, {Kroener}, {Kuindersma}, {Labadie}, {Leinert}, {Le Poole}, {Lizon},
  {Lucuix}, {Marcotto}, {Martinache}, {Martinot-Lagarde}, {Mathar}, {Matter},
  {Mauclert}, {Mehrgan}, {Meilland}, {Meisenheimer}, {Meisner}, {Mellein},
  {Menardi}, {Menut}, {Merand}, {Morel}, {Mosoni}, {Navarro}, {Nussbaum},
  {Ottogalli}, {Palsa}, {Panduro}, {Pantin}, {Parra}, {Percheron}, {Duc},
  {Pott}, {Pozna}, {Przygodda}, {Rabbia}, {Richichi}, {Rigal}, {Roelfsema},
  {Rupprecht}, {Schertl}, {Schmidt}, {Schuhler}, {Schuil}, {Spang},
  {Stegmeier}, {Thiam}, {Tromp}, {Vakili}, {Vannier}, {Wagner}, \&
  {Woillez}}]{Lopez14}
{Lopez}, B., {Lagarde}, S., {Jaffe}, W., {et~al.} 2014, The Messenger, 157, 5

\bibitem[{{Lucy}(1999)}]{Lucy99}
{Lucy}, L.~B. 1999, \aap, 344, 282

\bibitem[{{Luhman}(2018)}]{Luhman18}
{Luhman}, K.~L. 2018, \aj, 156, 271, \dodoi{10.3847/1538-3881/aae831}

\bibitem[{{Mannings} \& {Emerson}(1994)}]{Mannings94}
{Mannings}, V., \& {Emerson}, J.~P. 1994, \mnras, 267, 361,
  \dodoi{10.1093/mnras/267.2.361}

\bibitem[{{Marsh} \& {Mahoney}(1992)}]{Marsh92}
{Marsh}, K.~A., \& {Mahoney}, M.~J. 1992, \apjl, 395, L115,
  \dodoi{10.1086/186501}

\bibitem[{{McClure} {et~al.}(2013){McClure}, {D'Alessio}, {Calvet},
  {Espaillat}, {Hartmann}, {Sargent}, {Watson}, {Ingleby}, \&
  {Hern{\'a}ndez}}]{McClure13}
{McClure}, M.~K., {D'Alessio}, P., {Calvet}, N., {et~al.} 2013, \apj, 775, 114,
  \dodoi{10.1088/0004-637X/775/2/114}

\bibitem[{{Mendigut{\'\i}a} {et~al.}(2011){Mendigut{\'\i}a}, {Calvet},
  {Montesinos}, {Mora}, {Muzerolle}, {Eiroa}, {Oudmaijer}, \&
  {Mer{\'\i}n}}]{Mendigutia11}
{Mendigut{\'\i}a}, I., {Calvet}, N., {Montesinos}, B., {et~al.} 2011, \aap,
  535, A99, \dodoi{10.1051/0004-6361/201117444}

\bibitem[{{Mendoza V.}(1968)}]{Mendoza68}
{Mendoza V.}, E.~E. 1968, \apj, 151, 977, \dodoi{10.1086/149497}

\bibitem[{{Millan-Gabet} {et~al.}(1999){Millan-Gabet}, {Schloerb}, {Traub},
  {Malbet}, {Berger}, \& {Bregman}}]{Millan99}
{Millan-Gabet}, R., {Schloerb}, F.~P., {Traub}, W.~A., {et~al.} 1999, \apjl,
  513, L131, \dodoi{10.1086/311926}

\bibitem[{{Monnier} \& {Millan-Gabet}(2002)}]{Monnier02}
{Monnier}, J.~D., \& {Millan-Gabet}, R. 2002, \apj, 579, 694,
  \dodoi{10.1086/342917}

\bibitem[{{Monnier} {et~al.}(2005){Monnier}, {Millan-Gabet}, {Billmeier},
  {Akeson}, {Wallace}, {Berger}, {Calvet}, {D'Alessio}, {Danchi}, {Hartmann},
  {Hillenbrand}, {Kuchner}, {Rajagopal}, {Traub}, {Tuthill}, {Boden}, {Booth},
  {Colavita}, {Gathright}, {Hrynevych}, {Le Mignant}, {Ligon}, {Neyman},
  {Swain}, {Thompson}, {Vasisht}, {Wizinowich}, {Beichman}, {Beletic},
  {Creech-Eakman}, {Koresko}, {Sargent}, {Shao}, \& {van Belle}}]{Monnier05}
{Monnier}, J.~D., {Millan-Gabet}, R., {Billmeier}, R., {et~al.} 2005, \apj,
  624, 832, \dodoi{10.1086/429266}

\bibitem[{{Monnier} {et~al.}(2006){Monnier}, {Berger}, {Millan-Gabet}, {Traub},
  {Schloerb}, {Pedretti}, {Benisty}, {Carleton}, {Haguenauer}, {Kern},
  {Labeye}, {Lacasse}, {Malbet}, {Perraut}, {Pearlman}, \& {Zhao}}]{Monnier06}
{Monnier}, J.~D., {Berger}, J.-P., {Millan-Gabet}, R., {et~al.} 2006, \apj,
  647, 444, \dodoi{10.1086/505340}

\bibitem[{{Moshir} {et~al.}(1990){Moshir}, {Copan}, {Conrow}, {McCallon},
  {Hacking}, {Gregorich}, {Rohrbach}, {Melnyk}, {Rice}, {Fullmer}, \&
  {Chester}}]{Moshir90}
{Moshir}, M., {Copan}, G., {Conrow}, T., {et~al.} 1990, in IRAS Faint Source
  Catalogue, version 2.0 (1990)

\bibitem[{{Pauls} {et~al.}(2005){Pauls}, {Young}, {Cotton}, \&
  {Monnier}}]{Pauls05}
{Pauls}, T.~A., {Young}, J.~S., {Cotton}, W.~D., \& {Monnier}, J.~D. 2005,
  \pasp, 117, 1255, \dodoi{10.1086/444523}

\bibitem[{{Petrov} {et~al.}(2019){Petrov}, {Grankin}, {Gameiro}, {Artemenko},
  {Babina}, {Albuquerque}, {Djupvik}, {Gahm}, {Shenavrin}, {Irsmambetova},
  {Fernandez}, {Mkrtichian}, \& {Gorda}}]{Petrov19}
{Petrov}, P.~P., {Grankin}, K.~N., {Gameiro}, J.~F., {et~al.} 2019, \mnras,
  483, 132, \dodoi{10.1093/mnras/sty3066}

\bibitem[{{Pinilla} {et~al.}(2018){Pinilla}, {Tazzari}, {Pascucci}, {Youdin},
  {Garufi}, {Manara}, {Testi}, {van der Plas}, {Barenfeld}, {Canovas}, {Cox},
  {Hendler}, {P{\'e}rez}, \& {van der Marel}}]{Pinilla18}
{Pinilla}, P., {Tazzari}, M., {Pascucci}, I., {et~al.} 2018, \apj, 859, 32,
  \dodoi{10.3847/1538-4357/aabf94}

\bibitem[{{Pollack} {et~al.}(1994){Pollack}, {Hollenbach}, {Beckwith},
  {Simonelli}, {Roush}, \& {Fong}}]{Pollack94}
{Pollack}, J.~B., {Hollenbach}, D., {Beckwith}, S., {et~al.} 1994, \apj, 421,
  615, \dodoi{10.1086/173677}

\bibitem[{{Robitaille} {et~al.}(2007){Robitaille}, {Whitney}, {Indebetouw}, \&
  {Wood}}]{Robitaille07}
{Robitaille}, T.~P., {Whitney}, B.~A., {Indebetouw}, R., \& {Wood}, K. 2007,
  \apjs, 169, 328, \dodoi{10.1086/512039}

\bibitem[{{Schegerer} {et~al.}(2008){Schegerer}, {Wolf}, {Ratzka}, \&
  {Leinert}}]{Schegerer08}
{Schegerer}, A.~A., {Wolf}, S., {Ratzka}, T., \& {Leinert}, C. 2008, \aap, 478,
  779, \dodoi{10.1051/0004-6361:20077049}

\bibitem[{{Setterholm} {et~al.}(2018){Setterholm}, {Monnier}, {Davies},
  {Kreplin}, {Kraus}, {Baron}, {Aarnio}, {Berger}, {Calvet}, {Cur{\'e}},
  {Kanaan}, {Kloppenborg}, {Le Bouquin}, {Millan-Gabet}, {Rubinstein}, {Sitko},
  {Sturmann}, {ten Brummelaar}, \& {Touhami}}]{Setterholm18}
{Setterholm}, B.~R., {Monnier}, J.~D., {Davies}, C.~L., {et~al.} 2018, \apj,
  869, 164, \dodoi{10.3847/1538-4357/aaef2c}

\bibitem[{{Shakura} \& {Sunyaev}(1973)}]{Shakura73}
{Shakura}, N.~I., \& {Sunyaev}, R.~A. 1973, \aap, 24, 337

\bibitem[{{St-Onge} \& {Bastien}(2008)}]{StOnge08}
{St-Onge}, G., \& {Bastien}, P. 2008, \apj, 674, 1032, \dodoi{10.1086/524649}

\bibitem[{{STScI Development Team}(2013)}]{pysynphot}
{STScI Development Team}. 2013, {pysynphot: Synthetic photometry software
  package}, Astrophysics Source Code Library.
\newblock \doeprint{1303.023}

\bibitem[{{Takami} {et~al.}(2013){Takami}, {Karr}, {Hashimoto}, {Kim},
  {Wisniewski}, {Henning}, {Grady}, {Kandori}, {Hodapp}, {Kudo}, {Kusakabe},
  {Chou}, {Itoh}, {Momose}, {Mayama}, {Currie}, {Follette}, {Kwon}, {Abe},
  {Brandner}, {Brandt}, {Carson}, {Egner}, {Feldt}, {Guyon}, {Hayano},
  {Hayashi}, {Hayashi}, {Ishii}, {Iye}, {Janson}, {Knapp}, {Kuzuhara},
  {McElwain}, {Matsuo}, {Miyama}, {Morino}, {Moro-Martin}, {Nishimura}, {Pyo},
  {Serabyn}, {Suto}, {Suzuki}, {Takato}, {Terada}, {Thalmann}, {Tomono},
  {Turner}, {Watanabe}, {Yamada}, {Takami}, {Usuda}, \& {Tamura}}]{Takami13}
{Takami}, M., {Karr}, J.~L., {Hashimoto}, J., {et~al.} 2013, \apj, 772, 145,
  \dodoi{10.1088/0004-637X/772/2/145}

\bibitem[{{Tannirkulam} {et~al.}(2007){Tannirkulam}, {Harries}, \&
  {Monnier}}]{Tannirkulam07}
{Tannirkulam}, A., {Harries}, T.~J., \& {Monnier}, J.~D. 2007, \apj, 661, 374,
  \dodoi{10.1086/513265}

\bibitem[{{Tannirkulam} {et~al.}(2008){Tannirkulam}, {Monnier}, {Harries},
  {Millan-Gabet}, {Zhu}, {Pedretti}, {Ireland}, {Tuthill}, {ten Brummelaar},
  {McAlister}, {Farrington}, {Goldfinger}, {Sturmann}, {Sturmann}, \&
  {Turner}}]{Tannirkulam08}
{Tannirkulam}, A., {Monnier}, J.~D., {Harries}, T.~J., {et~al.} 2008, \apj,
  689, 513, \dodoi{10.1086/592346}

\bibitem[{{ten Brummelaar} {et~al.}(2005){ten Brummelaar}, {McAlister},
  {Ridgway}, {Bagnuolo}, {Turner}, {Sturmann}, {Sturmann}, {Berger}, {Ogden},
  {Cadman}, {Hartkopf}, {Hopper}, \& {Shure}}]{Brummelaar05}
{ten Brummelaar}, T.~A., {McAlister}, H.~A., {Ridgway}, S.~T., {et~al.} 2005,
  \apj, 628, 453, \dodoi{10.1086/430729}

\bibitem[{{ten Brummelaar} {et~al.}(2012){ten Brummelaar}, {Sturmann},
  {McAlister}, {Sturmann}, {Turner}, {Farrington}, {Schaefer}, {Goldfinger}, \&
  {Kloppenborg}}]{Brummelaar12}
{ten Brummelaar}, T.~A., {Sturmann}, J., {McAlister}, H.~A., {et~al.} 2012, in
  \procspie, Vol. 8445, Optical and Infrared Interferometry III, 84453C,
  \dodoi{10.1117/12.925023}

\bibitem[{{ten Brummelaar} {et~al.}(2013){ten Brummelaar}, {Sturmann},
  {Ridgway}, {Sturmann}, {Turner}, {McAlister}, {Farrington}, {Beckmann},
  {Weigelt}, \& {Shure}}]{Brummelaar13}
{ten Brummelaar}, T.~A., {Sturmann}, J., {Ridgway}, S.~T., {et~al.} 2013,
  Journal of Astronomical Instrumentation, 2, 1340004,
  \dodoi{10.1142/S2251171713400047}

\bibitem[{{Tuthill} {et~al.}(2001){Tuthill}, {Monnier}, \&
  {Danchi}}]{Tuthill01}
{Tuthill}, P.~G., {Monnier}, J.~D., \& {Danchi}, W.~C. 2001, \nat, 409, 1012,
  \dodoi{10.1038/35059014}

\bibitem[{{van der Marel} {et~al.}(2016){van der Marel}, {Verhaar}, {van
  Terwisga}, {Mer{\'{\i}}n}, {Herczeg}, {Ligterink}, \& {van
  Dishoeck}}]{vanderMarel16}
{van der Marel}, N., {Verhaar}, B.~W., {van Terwisga}, S., {et~al.} 2016, \aap,
  592, A126, \dodoi{10.1051/0004-6361/201628075}

\bibitem[{Van Der~Walt {et~al.}(2011)Van Der~Walt, Colbert, \&
  Varoquaux}]{van2011numpy}
Van Der~Walt, S., Colbert, S.~C., \& Varoquaux, G. 2011, Computing in Science
  \& Engineering, 13, 22

\bibitem[{{Zajtseva}(2010)}]{Zajtseva10}
{Zajtseva}, G.~V. 2010, Astrophysics, 53, 212,
  \dodoi{10.1007/s10511-010-9113-1}

\end{thebibliography}
\bibliographystyle{aasjournal}



\end{document}